\documentclass{article}

\usepackage{arxiv}

\usepackage[utf8]{inputenc} % allow utf-8 input
\usepackage[T1]{fontenc}    % use 8-bit T1 fonts
\usepackage{hyperref}       % hyperlinks
\usepackage{url}            % simple URL typesetting
\usepackage{booktabs}       % professional-quality tables
\usepackage{amsfonts}       % blackboard math symbols
\usepackage{nicefrac}       % compact symbols for 1/2, etc.
\usepackage{microtype}      % microtypography
\usepackage{lipsum}

\usepackage{graphicx}
\usepackage{tikz}
\usepackage{caption}
\usepackage{comment}
\usepackage{bm}
\usepackage{enumitem}
\usepackage{amsmath}
\def \nStates {12 }
\def \nhrus   {111}
\def \nLocalParAug {9 }
\def \nGlobalParAug{4 }

\def \change{black}
\usepackage{breqn}

\title{Daily Data Assimilation of a Hydrologic Model Using the Ensemble Kalman Filter}

\author{
  Sami A.~Malek \\
  UC Berkeley, Department of Electrical Engineering and Computer Sciences\\
  \texttt{sami.malek@berkeley.edu} \\
  \And
  Alexandre M.~Bayen \\
  UC Berkeley, Department of Electrical Engineering and Computer Sciences\\
  UC Berkeley, Department of Civil Engineering\\
  UC Berkeley, Institute for Transportation Studies\\
  \texttt{bayen@berkeley.edu} \\
  \And
  Steven D.~Glaser \\
  UC Berkeley, Department of Civil Engineering\\
  \texttt{bayen@berkeley.edu} \\
}

\begin{document}
\maketitle

\begin{abstract}
Accurate runoff forecasting is crucial for reservoir operators as it allows optimized water management, flood control and hydropower generation. Land surface models in mountainous regions depend on climatic inputs such as precipitation, temperature and solar radiation to model the water and energy dynamics and produce runoff as output. With the rapid development of cheap electronics applied in various systems, such as Wireless Sensor Networks (WSNs), satellite and airborne technologies, the prospect of practically measuring spatial Snow Water Equivalent in a dense temporal scale is increasing. We present a framework for updating the Precipitation Runoff Modeling System (PRMS) with Snow Water Equivalent (SWE) maps and runoff measurements on a daily timescale based on the Ensemble Kalman Filter (ENKF). Results show that by assimilating SWE daily, the modeled SWE gets updated accordingly, however no improvement is observed at the runoff model output. Instead, a deterioration consistently occurs. Augmenting the state space with model parameters and runoff model output allows for filter update with previous day measured runoff using the joint state-parameter method, and showed a considerable improvement in the daily runoff output of up to 60\% reduction in RMSE for the wet water year 2011 relative to the no assimilation scenario, and improvement of up to 28\% compared to a naive autoregressive AR(1) filter. Additional simulation years showed consistent improvement compared to no assimilation, but varied relative to the previous day autoregressive forecast during the dry year 2014.
\end{abstract}

% keywords can be removed

\section{Introduction}
\label{sec:intro}

%context
This research work fits into a larger objective of attaining consistently more accurate snowmelt runoff forecast resilient to climate change and extreme weather conditions. Presently, reservoir and dam operators rely on forecasts largely based on historical methods. Numerous studies show that climate change effects are challenging the stationarity assumption of this statistical approach. The past can no longer predict the future as well as it did before.  Other studies show a systematic receding date for snow melt during the year. \cite{CaDWR2017, bales06, Stewart2004, stewart05changes}

%flood control
\par A mandatory evacuation order was issued in February 2017 after the Oroville dam crisis because of hazardous water overflow at the Oroville reservoir. The Department of Water Resources (DWR) published a report showing that reservoir water levels rose earlier and higher than previous 16 years. Oroville is the pouring point of the North Fork Feather River where 3 days earlier, recently deployed sensors showed an onset of rapid snow melt up to 1 cm per hour most probably due to a rain on snow event. Unlike statistical models, a physically based forecast model that incorporates these new data should be able to capture such inflow surges. 

%reservoir management
\par In addition to flood control, runoff forecasting allows for better reservoir management and potentially more efficient power generation. Stakeholders, such as Pacific Gas \& Electric (PG\&E) have already started shifting towards physically based forecast models. These models though less vulnerable to climate change, require dense and accurate data as well as a methodology to seamlessly integrate sensor measurements of different temporal and spatial scales.

%newly deployed systems
\par Recent efforts to increase hydrologic instrumentation in the East Branch of the North Fork Feather River basin by a partnership between the University of California at Berkeley, the Department of Water Resources and the California Energy Commission (CEC) led to the deployment of four 1 km$^{2}$ scale state-of-the-art remote autonomous ground measurement systems based on Wireless Sensor Networks technology to collect temperature, humidity, solar radiation, snowdepth, soil temperature and soil moisture measurements at a 15 minute scale. These new deployments would supplement the existing measurement system consisting of 8 snow pillows reporting daily SWE measurements and the monthly manual snow course locations. Moreover the NASA airborne snow observatory, in partnership with DWR are using airborne LiDAR system to produce unprecedented bi-weekly basin-wide SWE maps of the Tuolumne and Merced river basins at 50 meter spatial resolution \cite{painter2016}.

%objective
\par The objective of this study is to design a practical data assimilation framework for real-time runoff forecasting using  PRMS that can leverage sensor data from the existing, state-of-the-art and future hydrologic measurement systems at different temporal and spatial scales.

\par
After describing the PRMS model dynamics, we present existing filtering techniques, then we develop a practical data assimilation framework to integrate in real-time simulated measurements of SWE and real measurements of streamflow in the model using the Ensemble Kalman Filter.

\label{chap:5_system_dynamics}
\section{Precipitation Runoff Modeling System}
\begin{figure}
\begin{center}
\includegraphics[scale=.7]{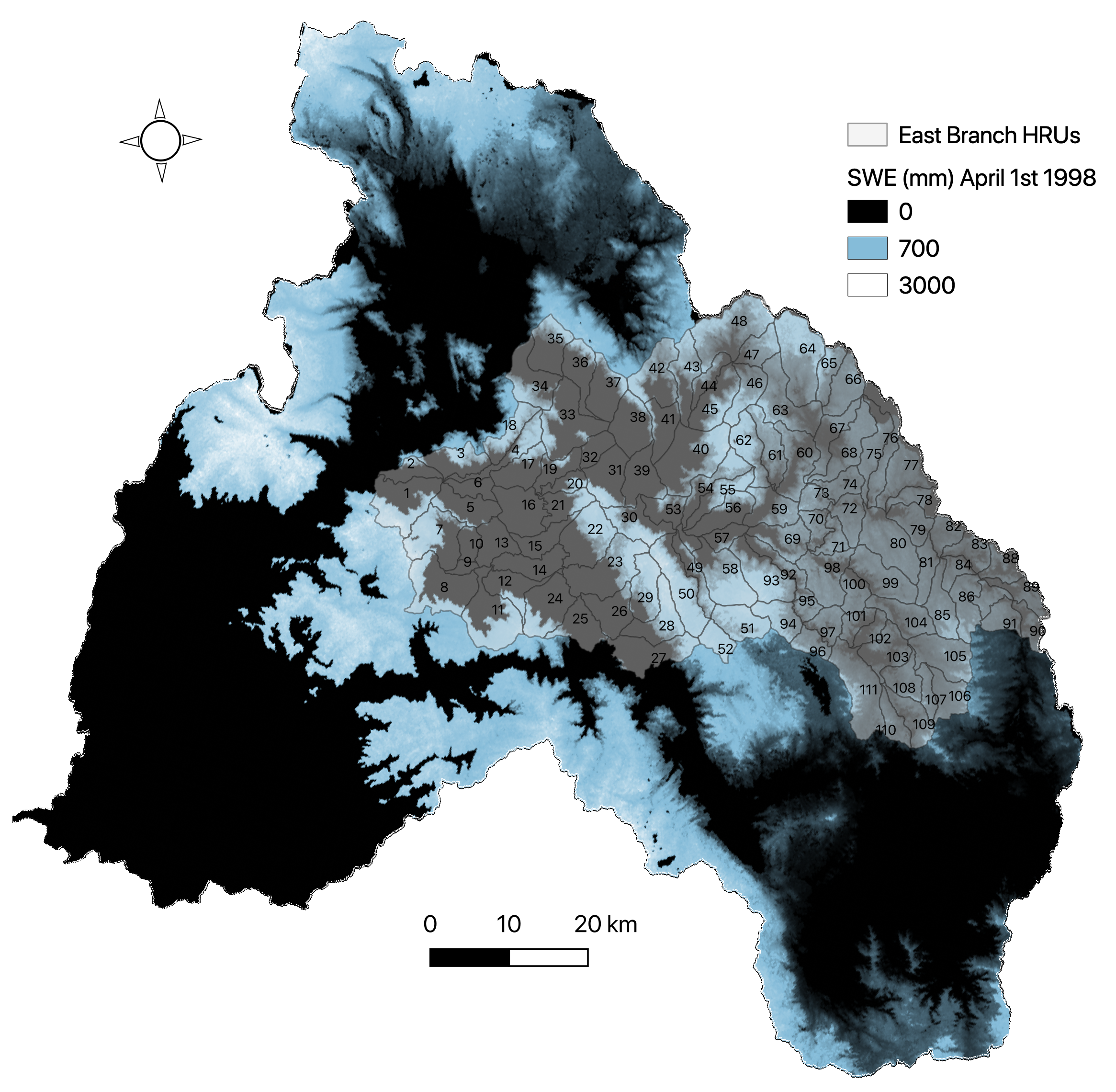}
\caption{\footnotesize{Hydrologic Response Units (HRUs) of the East Branch of the Feather River overlaid on a SWE map from 1998.}}
\label{fig:feather_hrus}
\end{center}
\end{figure}

%summary
The Precipitation Runoff Modeling System is a hydrologic model originally described by Leavesley (1983). 
It simulates snow accumulation and ablation processes as well as soil-zone and streamflow water dynamics.
%hrus
The watershed or basin is discretized into Hydrologic Response Units (HRUs) based on topography, land use, climate, soil properties, and geologic units information \cite{prms_calib}. Sample HRUs for the East Branch of the Feather River basin are shown in Fig. \ref{fig:feather_hrus}.

\begin{figure}
\begin{center}
\includegraphics[scale=.5]{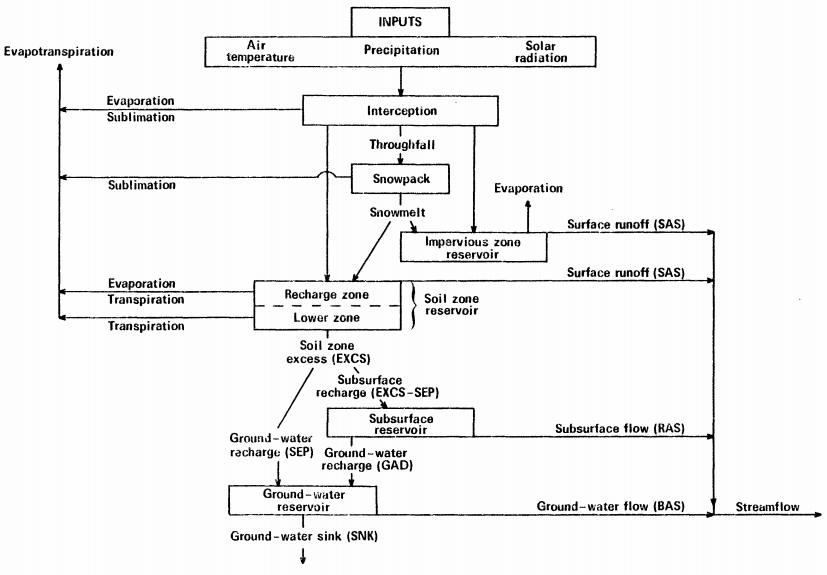}
\caption{\footnotesize{Schematic diagram of PRMS and its inputs \cite{prms}}}
\label{fig:prms_diagram}
\end{center}
\end{figure}

%timestep
The model runs on a daily time discretization, with some processes having half-day resolution.
%inputs
The model takes daily precipitation, minimum and maximum air temperature and optionally short-wave solar radiation as inputs. 
PRMS is a module based system: it can handle other data inputs as well as different modeling modules. We restrict this study to what is currently being operationally used in our specific application.
%In the following part, we will use consistent variable names as those used in the model with \textbf{bold} formatting indicating a model parameter. ??
% model states as water & energy containers 
Each HRU consists of a system of abstract water containers (or storages) interconnected by water flows as shown in Fig. \ref{fig:prms_diagram}. Variables with subscripts ``h" in this report indicate that those variables are unique for each HRU. 

The components are Interception ($Sint_{h}$), Snowpack ($SWE_{h}$ and $H_{h}$), Impervious Zone Reservoir ($Simp_{h}$), Soil Zone Reservoir ($Ssz_{h}$, subdivided into Recharge $Ssre_{h}$ and Lower zone $Ssl_{h}$), Subsurface Reservoir ($Sss_{h}$) and the Ground-Water Reservoir ($Sgw_{h}$). 
For each water storage, the water mass balance equation is satisfied.
\begin{equation}
\frac{\Delta\left(S\right)}{\Delta t} = \sum{}{}{F}
\end{equation}
where $S$ and $F$ denote water storage and flow respectively.
%%% interception
\begin{equation}
\frac{\Delta\left(Sint\right)}{\Delta t} =  \sum{}{}{
Fint_{in}} -  \sum{}{}{Fint_{out}}
\end{equation}
where $Fint_{in}$ consists of precipitation, and $Fint_{out}$ consists of evaporation, sublimation and througfall.
%, toSoilZone - toReservoir

%%%swe
\begin{equation}
\frac{\Delta \left(SWE\right)}{\Delta t} = \sum{}{}{
Fswe_{in}} -  \sum{}{}{Fswe_{out}}
\end{equation}
where $Fswe_{in}$ consists of precipitation on non-vegetated areas, throughfall from canopy covered regions, and $Fswe_{out}$ consists of sublimation and snowmelt.

%%%soil zone
\begin{equation}
\frac{\Delta\left(Ssz\right)}{\Delta t} = \sum{}{}{
Fsz_{in}} -  \sum{}{}{Fsz_{out}}
\end{equation}
where $Fsz_{in}$ consists of snowmelt, rain, and flow from interception on pervious regions, and $Fsz_{out}$ consists of surface runoff, evapotranspiration and excess $Sz$ water.

%%% ss
\begin{equation}
\frac{\Delta\left(Sss\right)}{\Delta t} =  \sum{}{}{
Fss_{in}} -  \sum{}{}{Fss_{out}}
\end{equation}
where $Fss_{in}$ consists of recharge from $Ssz$, subsurface flow to streamflow, and percolation to the ground water reservoirs ($Sgw$).

\begin{equation}
\frac{\Delta\left(Sgw\right)}{\Delta t} =  \sum{}{}{
Fsgw_{in}} -  \sum{}{}{Fsgw_{out}}
\end{equation}
where $Fsgw_{in}$ consists of recharge from Soil Zone and from subsurface reservoirs and $Fsgw_{out}$ consists of ground water sink and groundwater flow to sreamflow output.

For the Snowpack storage($SWE$), melt occurs when the snowpack reaches isothermal conditions ($H$ = 0). 
An additional energy balance equation is satisfied that deals with the atmospheric and temperature gradient and energy fluxes ($f$) interactions and can be written as:
\begin{equation}
\frac{\Delta\left(H\right)}{\Delta t} = \sum{}{}{f}
\end{equation}

%inputs
\begin{table}[h]
\begin{center}
\begin{tabular}{c|c|c|c|c|l}
Symbol & PRMS symbol & Type & Quantity & Units & Description \\
\hline
P & hru\_ppt & daily & per HRU & inches & total precipitation \\
Tmax, i & tmax & daily & one & $^{o}$F & max temperature at HRU i \\
Tmin, i & tmin & daily & one & $^{o}$F & min temperature at HRU i \\ \hline
\end{tabular}
\caption{Model Inputs}
\label{tab:model_inputs}
\end{center}
\end{table}
%states{c|c|c|c|l}
%\begin{tabularx}{\columnwidth}{c|c|c|c|X}
\begin{table}[h]
\begin{center}
\begin{tabular}{c|c|c|c|l}
Symbol & PRMS symbol & Quantity & Units & Description \\
\hline
Sint & intcp\_stor & per HRU & inches & interception storage \\
Sliq & freeh2o & per HRU & inches & liquid water snowpack storage \\
$\rho$ & pk\_den & per HRU & g/cm$^{3}$ & snowpack average density \\
SWE & pkwater\_equiv & per HRU & inches & total water snowpack storage \\
Sice & pk\_ice & per HRU & inches & frozen water snowpack storage \\
D & pk\_depth & per HRU & inches & snowpack depth \\
H & pk\_def & per HRU & Langleys & snowpack heat deficit \\
Tpk & pk\_temp & per HRU & $^{o}$F & snowpack temperature \\
fsca & snowcov\_area & per HRU & - & snow cover area \\
Sgw & gwres\_stor & per GWR & inches & ground water storage \\
Simp & imperv\_stor & per HRU & inches & impervious storage \\
Ssre & soil\_rechr & per HRU & inches & soil recharge storage \\
Ssz & soil\_moist & per HRU & inches & total soil-zone storage (soil-moisture) \\
Sss & ssres\_stor & per SSR & inches & sub-surface storage \\
SWEmax & pst & per HRU & inches & tracks max SWE of pack for fsca albedo \\ \hline
%\end{tabularx}
\end{tabular}
\caption{Model States}
\label{tab:model_states}
\end{center}
\end{table}

\newcommand{\cmmnt}[1]{\ignorespaces}

%%%%%%%%%%%%%%%%%%%%%%%%%
% Model Inputs
%%%%%%%%%%%%%%%
\subsection{Model Inputs}
Model inputs consist of daily time-series of precipitation for each HRU and minimum and maximum air temperature for 1 HRU that contains a temperature station. The daily short-wave radiation ($Rsw$) is estimated internally by the model. They are detailed below:

\underline{Short-Wave Radiation:}
%solar
A constant solar table consisting of daily estimates of the potential (clear sky) short-wave solar radiation ($Rpsw$) for each HRU (they are derived from calculations of duration of sun exposure and latitude for each HRU and each day \cite{prms}). 
%(Meeus, 1999). 
Such values are then adjusted daily based on a modification of the ``degree-day" method to get the short-wave radiation described in more details in \cite{prms}.
%by Leaf and Brink (1973) to get the short-wave radiation. It 
The method consists of deriving a degree-day coefficient ($dd$) and then a ratio of actual-to-potential radiation for a horizontal surface ($r$). %\cmmnt{sra2p}$solf_{HRU}$). 
The following parametric linear curve is used to find dd:
%dd coefficient
%\begin{equation}
%dd = ar \cdot T_{max} + br
%\end{equation}
%notation conforming
\begin{equation}
dd = \bm{\phi}_{m} \cdot Tmax_{f} + \bm{\beta}_{m}
\end{equation}
where $\bm{\phi}_{m}$ and $\bm{\beta}_{m}$ are parameters and $Tmax_{f}$ is an input. 
%where a and b are parameters, and $T_{max}$ is an input.
From dd, $r_{h}$ %\cmmnt{sra2p} $solf_{h}$ 
is retrieved via a non-linear relationship of $dd$ in function of  $r_{h}$. %\cmmnt{sra2p} $solf_{h}$. \todo{cite}
Finally the actual solar radiation is adjusted according to each HRU slope.
%swrad HRU
%\begin{equation} \label{eq_rad}
%swrad\_h = \frac{sra2p\_h\cdot radj}{cos\left(arctan\left(slp\_h\right)\right)} \cdot stps\_h 
%\end{equation}
%notation conforming
\begin{equation} \label{eq:swrad}
Rsw_{h} = \frac{r_{h} \times \bm{\gamma}}{cos\left(arctan\left(\bm{\delta}_{h}\right)\right)} \cdot Rpsw_{h} %soltab\_potsw_{h} 
\end{equation}
where 
\begin{comment}
\begin{equation}
\gamma =  \begin{cases}
{\gamma}s & \text{during summer \& ppt greater than ppt\_radj}\\
{\gamma}w & \text{during summer \& ppt greater than ppt\_radj}\\
0	  & \text{\& ppt less than ppt\_radj}
\end{cases}
\end{equation}
because the solar table was computed for days without precipitation.
\end{comment}

%\todo{fix me}
\begin{equation*}
\bm{\gamma} =  \begin{cases}
 \bm{{\gamma}s} &\text{during summer days \& $P$ greater than $\bm{Pmin}$}\\
 \bm{{\gamma}w} &\text{during winter days \& $P$ greater than $\bm{Pmin}$}\\
%${\gamma}s$ & \text{during summer days \& $P$ greater than $\bm{Pmin}$}
1 &\text{$P$ less than $\bm{Pmin}$}
\end{cases}
\end{equation*}
because $Rpsw_{h}$ %$soltab\_potsw_{h}$ 
is calculated for days without precipitation.

\underline{Air Temperature}
%tmax, tmin

The single temperature input at HRU i with elevation $e_{i}$ is spatially distributed for the remaining HRUs according to the following equation (and similarly for $T_{min}$) in degrees Fahrenheit.
\begin{equation}
Tmax_{h} = \bm{Tmax}_{h_{i}} - \bm{{\lambda}1}_{month} \times 
\left(\frac{\bm{e}_{h} - \bm{e}_{h_{i}}}{1000}\right) - \bm{{\beta}1}_{h}
\end{equation}

\begin{equation}
Tmin_{h} = \bm{Tmin}_{h{i}} -\bm{{\lambda}2}_{month} \times 
\left(\frac{\bm{e}_{h} - \bm{e}_{h_{i}}}{1000}\right) - \bm{{\beta}2}_{h}
\end{equation}

\underline{Precipitation}
%ppt
Precipitation $\mathit{P}_{h}$ for each HRU  are inputs to the model. 
The precipitation phase is determined solely based on temperature. 
If the HRU maximum air temperature ($Tmax_{h}$) is less than or equal to the parameter $\bm{Ts}$, all precipitation is snow. If $Tmin_{h}$ and $Tmax_{h}$ are greater than or equal to $\bm{Ts}$ and $\bm{Tr}$ respectively, precipitation is assumed to be all rain.
Otherwise, $P$ is considered a mixture of rain and snow.
The rain fraction of $P$ is computed as:
\begin{equation}
\label{eq:snow_rain}
fr_{h} = \left(\frac{Tmax_{h} - \bm{Ts}}{Tmax_{h} - Tmin_{h}}\right)\cdot \bm{\zeta_{m}}
\end{equation}
A value greater than 1 is considered an all rain event.
%%%%%%%%%%%%%%%%%%%%%%%%%
% Interception
%%%%%%%%%%%%%%%
\subsection{Interception}
Portions of the HRU with canopy cover intercept precipitation. 
Different canopy cover density values are used for the summer ($\bm{ds}_{h}$) and winter ($\bm{dw}_{h}$).  
Only when the precipitation amount exceeds the canopy storage capacity ($Sc$) does it reach the ground as throughfall ($Pt_{h}$).
For example the net precipitation reaching the ground during summer is:
%\begin{equation}
%P\_n_{h}	= thf_{h} \cdot ccovs + \left(1.0 - ccovs\right) \cdot ppt_{h} 
%\end{equation}
\begin{equation}
\label{eq:net_ppt}
Pn_{h}	= Pt_{h} \cdot \bm{ds}_{h} + \left(1.0 -  \bm{ds}_{h}\right) \times \mathit{P}_{h} 
\end{equation}
where the throughfall $Pt_{h}$ is computed as:

\begin{equation}
Pt_{h} = \begin{cases}
P_{h} - \frac{Sc_{h}}{\bm{A}_{h} \cdot  \bm{ds}_{h}} & :P_{h} > \frac{Sc_{h}}{\bm{A}_{h} \cdot \bm{ds}_{h}}\\
0.0 & :\text{otherwise}
\end{cases}
\end{equation}
where the canopy storage capacity $Sc_{h}$ is:
%\begin{equation}
%cStr_{h} = \left(r\_itcps - itcpStr_{h}\right) \cdot A_{h} \cdot ccS_{h}
%\end{equation}
\begin{equation}
Sc_{h} = \left(\bm{Crs}_{h} - Sint_{h}\right) \cdot \bm{A}_{h} \cdot \bm{ds}_{h}
\end{equation}
Parameters $\bm{ds}_{h}$ and $\bm{Crw}_{h}$ are used during winter. A similar process is modeled for snow. 
$Sint_{h}$ is a model state holding intercepted precipitation. 
$\bm{Crs}_{h}$ and $\bm{Crw}_{h}$ are the summer and winter rain interception storage capacity for the major vegetation type for each HRU. $\bm{Cs}_{h}$ is used for snow.
%Parameters $ccw_{h}$ and $r\_itcpW$ are used during winter. A similar process is modeled for snow.
%itcp\_str_{h} is a model state holding intercepted precipitation. 
Intercepted rain, (or snow) evaporates (or sublimates) at a free-water surface rate.
%(at a fraction of the parameter potential evapo-transpiration: pet\_sub).

%%%%%%%%%%%%%%%%%%%%%%%%%
% Evapo-transpiration
%%%%%%%%%%%%%%%
\subsection{Evapotranspiration}
\label{ssec:et}
The Jensen-Haise formulation \cite{prms} is used to compute the potential evapotranspiration:

\begin{equation}
ET_{h} = \bm{jc} \cdot \left(T_{h} - jc_{h}\right)\cdot \frac{Rsw_{h}}{2.54\cdot \left(597.3 - \left(0.5653 \cdot T_{h}\right)\right)}
\end{equation}
where the coefficient $jc_{h}$ is approximated to be:
\begin{equation}
jc_{h} = 22 - \frac{\bm{e_h}}{1000}
\end{equation}
As figure \ref{fig:prms_diagram} shows, evapotranspiration occurs in multiple reservoirs.

%%%%%%%%%%%%%%%%%%%%%%%%%
% Snowpack
%%%%%%%%%%%%%%%
\subsection{Snowpack}
% Snow dynamics
Snowpack dynamics are modeled using water and energy balances for each HRU. 
The snowpack is abstracted as 2 layers. 
The energy exchanges that occur between the snowpack and the atmosphere are radiative, conductive and convective.
% energy balance
The energy balance at the snow-atmosphere interface is computed twice per day, for both the day and night periods as:
\begin{equation} \label{eq_eb}
\frac{dE}{dt} = I_{h} + Qv_{h} + R_{h}	%cal + cecsub + swn
\end{equation}
\begin{description}[align=right,labelwidth=1cm, itemsep=0mm]
\item[$E$] is the surface energy of the snowpack
\end{description}
The net short wave radiation $R_{h}$ in (\ref{eq_eb}) is computed from $Rsw_{h}$ (\ref{eq:swrad}) after accounting for the reflected portion of the radiation using the surface albedo $\alpha$ and that limited by the vegetative transmission coefficient parameter $\bm{\psi_{h}}$.
\begin{equation}
R_{h} = \left(1 - \alpha_{h}\right) \cdot \bm{\psi_{h}} \cdot Rsw_{h} 
\end{equation}
Where $\alpha$ is obtained from a non-linear curve that is a function of the snow dynamics phase and time since last new snow in days.

The net incoming longwave radiation $I_{h}$ originates from the atmosphere and land cover:
\begin{equation}
I_{h} = \bm{cw_{h}} \cdot Ip_{h} + \left(1-\bm{cw_{h}}\right)\cdot \bm{\epsilon}\cdot Ip_{h}
\end{equation}
where $\bm{\epsilon}$ is the emissivity, a precipitation situation-dependent parameter (between 0.0 and 1.0 dependent on storm type), and $Ip_{h}$ is the perfect black-body emission:
\begin{equation}
\label{eq:lwp_power4}
Ip_{h} = 5.85 \cdot 10^{-8}\cdot \left(T_{h} + 273.16\right)^{4}
\end{equation}
The combined convection and latent heat flux from condensation is modeled as linear function of temperature.
\begin{equation}
Qv_{h} = \bm{\omega} \cdot T_{h}
\end{equation}
When (\ref{eq_eb}) is negative, heat flow occurs by conduction only (ex: potentially refreeze snow at isothermal) and depends on the snowpack density and layer temperature gradient:
\begin{equation}
Qc_{h} = 2 \cdot \left[ 0.5 \cdot \sigma \sqrt{\frac{0.0077 \cdot \rho_{h}^{2} \cdot \Delta t}{ \sigma \cdot \rho_{h}}} \right] \cdot \left(T_{h} - Tpk_{h}\right)
\end{equation}
where $\sigma$ is the specific heat of ice 
%(in calories per cubic centimeter per degrees 
(in Celsius $Cal.m^{-3}.^{o}C^{-1}$).
A positive energy exchange translates into snowmelt at the surface of the snowpack, that transports heat $\Delta E$ into the lower snowpack by mass transfer. 
The potential melt from such heat is:
\begin{equation}
M_{h} = \frac{\Delta E}{203.2} \cdot sca_{h}
\end{equation}
where $sca_{h} = fsca_{h} \cdot \bm{A}_{h}$ is the HRU snow covered area.

The energy state of the lower layer is represented as a heat deficit. 
The snowpack heat deficit ($H$) for each HRU represents the amount of heat necessary to bring the snowpack to isothermal 0 $^o$C. 
Snowmelt occurs only when the heat deficit reaches zero and the snowpack free-water storage is exceeded.
Otherwise, potential melt is ``refrozen" by the decrease in the heat deficit.
%It is computed directly from the snowpack's negative temperature (Tpk) and SWE.
%\begin{equation}
%H_{h} = -Tpk_{h} \cdot SWE_{h} \cdot 1.27
%\end{equation}
%where 1.27 is the specific heat of ice in calories per inch per square cm per degree Celsius; SWE is the pack water equivalent in inches. 
Storage within the snowpack is tracked in two states: ice (solid) and free water (liquid), the sum of which is termed Snow Water Equivalent ($SWE$), and quantifies the volume of water obtained if all snow is melted.
Melt decreases the ice storage of the snowpack, and once the amount of free water surpasses the maximum capacity, water exits the snowpack.
%sublimation 
Sublimation loss from the snowpack occurs according to the following equation:
\begin{equation}
B_{h} = \bm{\xi} \cdot ET_{h} \cdot fsca_{h}
%\right) - subi_{h}
\end{equation}
where $ET_{h}$ is the potential evapotranspiration described in \ref{ssec:et} %and $subi_{h}$ is the sublimation from intercepted snow. 
The process lowers the non-isothermal heat deficit by: $Tpk_{h} \cdot ET_{h} \cdot 1.27$ calories.

% snowdepth
The evolution of the snow depth ($D$) is approximated from the ordinary differential equation \cite{prms}:
%(Riley and others, 1973) \todo{cite}:
\begin{equation}
\frac{dD_{h}}{dt} + \bm{\tau} \cdot D_{h} = \frac{Pns_{h}}{\bm{\rho_{init}}}+\left(\bm{\tau} \cdot \frac{SWE_{h} + Pns_{h}}{\bm{\rho_{max}}}\right)
\end{equation}
and is calculated as:
\begin{equation}
\Delta D_{h} = \Delta t \cdot \left(\frac{Pns_{h}}{\bm{\rho_{init}}} + \bm{\tau} \cdot \left(\frac{SWE_{h} + Pns_{h}}{\bm{\rho_{max}}} - D_{h}\right)\right)
\end{equation}
Where $\bm{\rho_{init}}$ and $\bm{\rho_{max}}$ are parameters for initial and maximum density of new snowfall, respectively.
$Pns_{h}$ is the net new snowfall after interception.
The snowpack density is then:
\begin{equation}
\rho_{h} = \frac{SWE_{h}}{D_{h}}
\end{equation}

The snow-covered area is determined by a multi-modal depletion curve \cite{prms}
%\todo{cite (Anderson, 1973).} 
that describes the evolution of the fractional snow-covered area ($fsca$) in function of the fraction of maximum $SWE$. 
Parameter $\bm{Acurve}$ specifies the 11 values of the curve. The depletion curve is used when HRU $SWE$ is less than the maximum $SWE$ corresponding to total snow cover specified by parameter $\bm{SWEmax}$.

Water available at the soil surface (from melt and rain) proceed to both infiltrating the soil reservoir and filling the impervious zone reservoir. 
%In both cases, once surface water overfills the capacity of these reservoirs, it becomes surface runoff.

%%%%%%%%%%%%%%%%%%%%%%%%%
% Impervious
%%%%%%%%%%%%%%%
\subsection{Impervious Zone Reservoir}
%impervious zone reservoir
The impervious zone reservoir ($Simp$) constitutes the fraction of the HRU specified by the parameter $\bm{fi}$ with no soil-infiltration capacity. It receives $\bm{f_{i}}$ fraction of the total water from snow melt ($M$) and rain throughfall ($Pnr$) for each HRU and has maximum retention capacity of $\bm{Simax}$, above which water flows directly as surface runoff $Fsri$ described in \ref{ssec:sroff}.
The process is described in the following discretized dynamic equation:
\begin{equation}
\label{eq:impStr}
Simp = Simp_{t-1} + (Pnr_{h} + M_{h})\cdot \bm{fi_{h}} - Fsri_{h} - ei_{h}
\end{equation}
where ei is the water depleted by evaporation:
\begin{equation}
ei_{h} = ETa_{h} \cdot(1-fsca_{h}) \cdot \bm{fi_{h}}
\end{equation}
where $ETa_{h}$ is the available potential for evapotranspiration remaining in the system after accounting for evapotranspiration and sublimation losses that already occured (interception,..).
\subsection{Soil Zone Reservoir}
%soil zone reservoir
The soil zone reservoir ($Ssz$) represents the active soil profile and its depth is considered to be the average rooting depth of the predominant vegetation on the HRU. 
It receives part of the water from snow melt and rain. 
It can hold up to the parameter $\bm{Sszmax}$. 
The upper layer is termed the recharge zone ($Ssre$) and the lower layer is termed the lower zone ($Ssl$). 
While water is lost through both evaporation and transpiration from the upper layer, it is only lost through transpiration from the lower layer. Evaporation is limited by both how much potential of $ET$ is left in $ETa_{h}$ for the day and by the availability of water.
Water needs to fill the recharge layer before it can proceed to the lower layer.
The soil water content represented by $Ssz$ is also defined as the ``soil moisture". 
A maximum of $\bm{Fzgwmax_{h}}$ percolates from $Ssz$ to $Sgw$, and the remainder to $Sss$ as excess.
%%%%%%%%%%%%%%%%%%%%%%%%%
% Surface runoff
%%%%%%%%%%%%%%%
\subsection{Surface Runoff}
\label{ssec:sroff}
Both the pervious soil zone and the impervious reservoirs contribute to surface runoff $Fsr$ as $Fsrp$ and $Fsri$ respectively. %Water above the soil zone capacity needs to also overcome the snow infiltration rate.
\begin{equation}
Fsr_{h} = (1-\bm{fi_{h}}) \cdot Fsrp_{h} + \bm{fi_{h}} \cdot Fsri_{h}
\end{equation}
The pervious runoff is modeled using the following non-linear parametric equation in function of soil moisture, net rain precipitation and snowmelt \cite{prms}:
%inspired from \todo{Dickinson and Whiteley (1970)}:
\begin{equation}
\label{eq:Fsr_exp}
Fsrp_{h} = max\left[\bm{{\alpha}1_{h}} \cdot 10^{\bm{{\theta}1_{h}} \cdot Ssz \cdot (0.5 \cdot Pn_{h})}, \bm{Asr}\right] \cdot \left(Pnr_{h} + M_{h}\right)
\end{equation}
Where $Pn_{h}$ is net precipitation (\ref{eq:net_ppt}), $\bm{{\alpha}1_{h}}$, $\bm{{\theta}1_{h}}$ and $\bm{Asr}$ are model parameters.
The impervious runoff is equal to the amount of water exceeding $\bm{Simax}$ in (\ref{eq:impStr}).

%%%%%%%%%%%%%%%%%%%%%%%%%
% ss
%%%%%%%%%%%%%%%
\subsection{Subsurface Reservoir}
The subsurface reservoir models the relatively rapid water movement of the unsaturated soil zones towards a streamflow channel. 
Such behavior is highlighted typically during and after a rainfall or snowmelt event.
A reservoir routing system is used to model the flow from the subsurface reservoir. 
The flow $Fss$ is the solution that satisfies both the mass continuity equation:
\begin{equation}
Fss_{s, t} = Fss_{s, t-1} -  \frac{d\left(Sss\right)}{dt}
\end{equation}
and the empirical quadratic relation:
\begin{equation}
\label{eq:Fss_square}
Fss_{s, t} = \bm{\alpha}3_{s} \cdot Sss_{s} + \bm{\beta}3_{s} \cdot Sss^{2}_{s}
\end{equation}
where $\bm{\alpha2}_{s}$ and $\bm{ \bm{\beta}2_{s}}$ are routing coefficients.

Another $Sss$ discharge route transfers water to the $Sgw$ reservoir through the parametric power equation:
\begin{equation}
\label{eq:Fss_exp}
Fss\_gw_{s} = \bm{\alpha2_{s}} \cdot \left[\frac{Sss_{s}}{\bm{smax_{s}}}\right]^{\bm{\theta2_{s}}}
\end{equation}
where $\bm{\alpha2_{s}}$, $\bm{smax}$ and $\bm{\theta2_{s}}$ are model parameters.
%Ssr2gw_rate, Ssrmax_coef and Ssr2gw_exp respectively.

%%%%%%%%%%%%%%%%%%%%%%%%%
% gw
%%%%%%%%%%%%%%%
\subsection{Ground-Water Reservoir}
Once the soil zone reservoir is saturated, excess water starts filling the ground-water ($Sgw$) reservoir at a maximum daily recharge rate $\bm{Fzgwmax_{h}}$, a parameter defined for each HRU. 
\begin{equation}
Fsz\_gw = min(\bm{Fzgwmax_{h}}, Ssz\_excess)
%\bm{grr_{h}} \cdot fp_{h} 
\end{equation}
%\todo{check below claim, unlikely}
%and at a rate of $\bm{ssg}$ from the subsurface reservoirs (Sss):
%\begin{equation}
%Fss\_gw = \bm{ssg_{h}} \cdot Ass_{h}
%\end{equation}
Water leaving the ground-water reservoir either flows laterally ($Fgw$) contributing to the baseflow portion of the streamflow or is lost from the system via the groundwater sink, both at a linear rate:
\begin{equation}
Fgw_{h} = \bm{\alpha4_{g}} \cdot Sgw_{g}
\end{equation}
\begin{equation}
Fgsnk_{h} = \bm{\alpha5_{g}} \cdot Sgw_{g}
\end{equation}
%Note that gw and ss reservoirs need not be aligned with the HRU and could be recharged from multiple HRUs, however 

%%%%%%%%%%%%%%%%%%%%%%%%%
% output
%%%%%%%%%%%%%%%
\subsection{Model Output: Streamflow}
Basin streamflow, which consititues the model output, is computed as an area weighted average of all the reservoir flow outputs described above:
\begin{equation}
\label{eq:F_basin}
F_{basin} = \sum_{h = 1}^{N_{h}} \frac{Fsr_{h}}{A_{h}} +  \sum_{s = 1}^{N_{s}} \frac{Fss_{s}}{A_{s}} +
\sum_{g = 1}^{N_{g}} \frac{Fgw_{g}}{A_{g}} 
\end{equation}
Improving the daily forecast accuracy of $F_{basin}$ is the ultimate objective of this report.

%%%%%%%%%%%%%%%%%%%%%%%%%%
%   Filtering Techniques
%%%%%%%%%%%%%%%%%%%%%%%%%

\section{Filtering Techniques}

%error modeling
\def \stdPm{0.4 } 	%
\def \stdT{2 } 		%celsius
\def \stdSWEm{0.1 }	
\def \stdRoffm{0.005}
\color{\change} \ 
Models are simplifications of physical processes and are inherently imperfect representations of reality. Combining related measurements with modeled estimates through data assimilation would produce more accurate outcomes. Assimilation approaches were introduced in oceanography and meteorology \cite{oceo1, oceo2} with different schemes used such as the variational methods \cite{variational} and the filtering techniques including nudging \cite{nudging}, optimal interpolation \cite{solaris} and Kalman filtering with its variants that are presented below. 
\color{black}

\underline{Bayesian Filtering}

% Bayesian filtering
Bayesian inference \cite{bayes} is the underlining principle for all the data assimilation approaches we will describe in this section.
They stem from the two following probability rules: marginalization (sum) rule:
Given $f_{xy}(x, y)$, we define: %\todo{integral instead}
\begin{equation}
f_{x}(x) := \sum_{y \epsilon Y} f_{xy}(x, y)
\end{equation}
and the conditioning, or product rule defined as:
\begin{equation}
f_{x|y}(x, y) := \frac{f_{xy}(x, y)}{f_{y}(y)}
\end{equation}
for $f_{y}(y)\neq 0$
%total probability theorem

We can formulate the following forecast and update system in the Bayesian framework, where $x$ represents the model state, $z$ represent the measurements and $k$ is the discrete time step:

Prior update (or forecast):
\begin{equation}
\label{eq:prior_prime}
f(x(k)|z(1:k-1)) = \int f(x(k)|x(k-1)) \cdot f(x(k-1)|z(1:k-1)) dx(k-1)
\end{equation}
where the first item in the integral is the process model and the second item is the previous time step measurement update result.

Measurement update (or update):
\begin{equation}
\label{eq:updated_prime}
f(x(k)|z(1:k)) = \frac{f((z(k)|x(k))\cdot f(x(k)|z(1:k-1)}{\int f(z(k)|x(k)) f(x(k)|z(1:k-1))dx(k)}
\end{equation}
where the first and second items of the numerator are the measurement model and the prior (\ref{eq:prior_prime}), respectively, and the denominator is the normalization $f(z(k)|z(1:k-1))$.

%KF
\underline{Kalman Filter} 
It can be shown \cite{Kalman_Rudolf, kf-proof, kaipio_somersalo_2010} that given the system of equations with a linear process model A and independent Gaussian distributions for errors $v(k-1)$ and $w(k)$ and prior:
assuming initialization: $\hat{x}_{a}(0) = x_{0}, P_{a}(0)=P_{0}$
\begin{equation}
x(k) := A(k)\cdot x(k-1) + u(k-1) + v(k-1) \newline
\end{equation}
\begin{equation}
y(k) := H(k)x(k) + w(k)
\end{equation}
the analytical solution to the Bayesian state estimation problem in (\ref{eq:prior_prime}) and (\ref{eq:updated_prime}) is the Kalman Filter (KF) and can be written in a recursive form while keeping track of the distributions' mean and covariance as:
\newline Prior update/Forecast step:
\begin{equation}
\hat{x}_{f}(k) = A(k-1)\hat{x}_{a}(k-1) + u(k-1)
\end{equation}
\begin{equation}
P_{f}(k) = A(k-1)P_{a}(k-1)A^{T}(k-1)+Q(k-1)
\end{equation}
\newline A Posteriori update/Analysis step:
\begin{equation}
\label{eq:kf_update}
\hat{x}_{a}(k) = \hat{x}_{f}(k) + P_{a}(k)H^{T}(k)R^{-1}(k)\left(z(k) - H(k)\hat{x}_{f}(k)\right)
\end{equation}
\begin{equation}
P_{a}(k) = \left(P_{f}^{-1}(k) + H^{T}(k)R^{-1}(k)H(k)\right)^{-1}
\end{equation}
without the need to store all previous observations and states. 
It is commonly referred to $P_{a}(k)H^{T}(k)R^{-1}(k)$ in (\ref{eq:kf_update}) as the Kalman gain $K(k)$.
The maximum a posteriori estimate (MAP) of the 
%\todo{you sure did not copy/paste? paraphrase} 
%unimodal and symmetric 
normally distributed analysis coincides with the mean:
$x^{MAP}_{a}(k) = \hat{x}_{a}(k)$ and thus is chosen as the best estimate of the system state at time k.
  
  %Similarly, updating an observation grid cell results in the update of distant unknown grid cells if a spatial covariance between them exists. 
  The basic Kalman Filter (KF) algorithm that is restricted for only linear models consists of two steps, a predict step followed by an update step.\color{\change} \ The KF and its variants have numerous applications in the physical sciences, primarily in the guidance \cite{ballistic}, navigation and control of vehicles \cite{auto}, time-series analysis \cite{market} and robotics \cite{zarchan_musoff_2005}.
  \color{black}
  
  %EKF
  \underline{Extended Kalman Filter}
  
  The Extended Kalman Filter (EKF) is an extension to the KF for non-linear systems. The approach is to linearize the non-linear process model around the mean using first order Taylor expansion. It thus becomes an approximation of the exact solution. The Jacobian is then used to advance the error covariance matrix in time. It requires the model to be continuously differentiable and is shown to have bad performance when the system exhibits strong non-linearities.\color{\change} \ For instance, the EKF was successfully used to integrate GPS-derived flow velocity measurements from drifters to improve the non-linear hydrodynamics state estimates in a controlled channel pilot experiment \cite{extended_bayen}.\color{black}
  
  %UKF
  \underline{Unscented Kalman Filter}
  
  The Unscented Kalman Filter (UKF), also referred to as Sigma-Point KF, tries to solve the linearity problem using another approach. 
  It works by first deterministically sampling using an algorithm sigma points from the distribution and assigning to each point a weight. 
  At every time step the points are individually propagated through the non-linear model, and finally the resulting mean and covariance is computed from the forecasted sigma points. 
  Unlike the EKF, not only the mean is advanced in time, but also many points sampled from the distribution. The method requires N = 2r + 1 sigma points where r is the system dimension.\color{\change} \ \cite{ukf_vs_ekf} shows that the UKF, with the same complexity, is a better filter than the EKF for non-linear models capturing sigma points' mean and covariance accurately up to the third order (in Taylor series expansion terms) while the EKF only to the first. Using UKF instead of EKF halfed the RMSE in estimating vehicle and wheel angular velocity of the anti-lock braking system (ABS) in \cite{ukf_double_ekf}. \color{black}

\underline{Ensemble Kalman Filter}
 
 The EnKF represents the distribution as a random sample of points called ensemble members. 
 In the forecast step, each ensemble member is integrated forward in time with an additional process noise.
The forecasted error covariance matrix is approximated at any time step by the ensemble sample covariance. 
The error covariance matrix is thus implicitly propagated saving the high computational requirements associated with its storage 
%\todo{(UKF?)} 
and forward integration such as in KF and EKF. 
The update stage consists of updating every ensemble member using sample covariance as an approximation of the true covariance. 
The ensemble members are never resampled from the distribution thus preserving the ensemble skewness, kurtosis, clustering, etc.
In the EnKF update, the updated ensemble is obtained by shifting and re-scaling the forecast ensemble.

\underline{Particle Filter}

 The Sequential Importance Resampling (SIR) Particle Filter (PF),  samples points called particles from the distribution to generate system states realizations.
 These particles are assigned weights that represent their likelihood. 
 The sampling is done randomly like the EnKF, but not necessarily from normal distributions. 
 On update, the particle weights are adjusted. 
 Higher weights are assigned to those that are supported by sensor data and vice versa.  
 The Particle Filter suffers from a problem with sample degeneracy and impoverishment, in that some scenarios end up with very few high-weight particles while the other majority of particles have almost zero weights. 
Another disadvantage is that the number of particles required for good performance scales exponentially with the model dimension.
 The method does not require Gaussian distributions nor model linearity \cite{pf_tut}.
 It thus works well with distributions that have complex shape and multiple peaks. 
\color{\change}Example applications include vehicle localization \cite{pf_car}, indoor occupant positioning using a Radio Frequency Identity RFID and receivers \cite{PF_position}, etc. \color{black}
 %\todo{TODO: indoor occupant}
 %\todo{TODO: open channel flow x2}
 %For instance, it was preferred  

 %filter selection: reasons why you chose this, and why this is better
 \underline{Filter Selection}
 
 %KF
 The Kalman Filter in its basic form cannot be applied to our modeling system because the latter is non-linear.
 %EKF
 The model is also not continuously differentiable, which eliminates the option of using the Extended Kalman Filter. 
 For example, different physical conditions can lead to the use of totally different process equations. 
 %(Note that some work exist where EKF has been adapted to non-differentiable models by introducing modalities \todo{cite snow17 dissertation})
 %UKF
For our application, using either of UKF, EnKF or PF is theoretically valid. 
According to \cite{nSigma}, UKF would require $\the\numexpr((\nStates + \nLocalParAug)*\nhrus + \nGlobalParAug + 1)+1\relax$  sigma points to fully represent the mean and variance of the system.
In \cite{ukf_roff}, the UKF was used in a similar application as ours but using algorithms to reduce the number of sigma points to $n + 2$.
%EnKF
Multiple studies in the literature pertaining to geophysics, oceanography and hydrology have shown that the EnKF performs well even when the ensemble size is orders of magnitude smaller than the system dimension \cite{hess_enkf_p, hess-enkf_N_L}.
%for instance an ensemble size of $10^{2}$ and a dimension of $10^{6}$ \todo{cite}.
EnKF outperformed PF in the assimilation of a conceptual rainfall-runoff in \cite{pf_vs_enkf} and the justification was that it is ``less sensitive to misspecification of the model and uncertainties", and in \cite{pf_kf_2same}, it had similar performance.
Given the above specified reasons and the limitation in resources available that will restrict the number of model realizations, we choose the EnKF.

\subsection{EnKF \& State-space}
The Ensemble Kalman Filter \cite{Evensen2003} was introduced 
%by Evensen 
in \cite{eve} as an alternative to EKF to overcome specific difficulties with nonlinear state evolution models, including non-differentiability of the model and closure problems.
 As described above, EnKF uses Monte Carlo (or ensemble integrations) and has the same standard update equations of the KF, with state mean and covariance replaced by an estimated ensemble sample mean and covariance respectively \cite{Eve3}. 
\color{\change}
The EnKF and more generally statistical inversion schemes were greatly advanced in \cite{ks_book}, mainly in the context of Tomography.
A framework for sampling a smooth prior is presented, especially when structural information about the ensemble (correlations between pixels) is known apriori. Such techniques can also be used in the perturbations of spatial data such as climatologic inputs, model error etc. For instance these techniques were used to generate smooth initial spatial model states ensemble in \cite{mobileMillenium} without shocks. \cite{ks_cov} shows that using a non-trivial state covariance matrix in the state noise yields much better performance than a stationary one. 

With the Kalman Smoother, state estimation was applied in Electrical Impedance Tomography (EIT) to reconstruct rapidly time-varying cross-sectional body scans inferred from boundary voltage and current measurements \cite{ks_elec_imped}, \cite{ks_elec_imped2}. Similarly, Electrical Resistance Tomography (ERT) measurements were used to infer water saturation levels in soil using EKF assimilation of a weakly non-linear model \cite{kaipio_vadose}. Time-dependent noise was used in \cite{ks_timeseries}  infer gas temperature from resistance measurements.

\cite{mobileMillenium} provided a framework for assimilating readily available GPS measurements of vehicle speed and velocity into a traffic model \cite{traffic_model_da} using the EnKF as part of the the Mobile Millennium traffic-monitoring system\cite{millenium_sys}.
Results of an unprecedented scale experiment \cite{mobileMillenium}, where GPS speed and position data from 100-vehicles were assimilated into the traffic model, showed improved estimates of the traffic state even with very low percentage of GPS-equipped vehicles participating.
\color{black}

We now present the conceptual algorithm of the EnKF: 
 Let \textit{s} be the system dimension.
 %(includes model states \textit{m}).
 Hydrologic models typically take inputs. 
 Let \textit{i} be the number of inputs and \textit{o} be the number of observations. Let $\bm x$ be the state vector and $\mathcal{M}$ the non-linear model. The discrete time k forward integration equation can be represented as:
 \begin{equation}
 \bm x^{k} = \mathcal{M}(\bm x^{k-1}, \bm b^{k})
 \end{equation}
 Where the model observed states are:
 \begin{equation}
 \bm y^{k} = \bm H \bm x^{k}
 \end{equation}
 The localized EnKF assimilation scheme 
 %with online parameter update 
 can be summarized by the algorithm below:
 
\begin{enumerate}[leftmargin=3\parindent]
\item \textit{Initialization}: Draw N ensemble realizations $\bm x^{0}_{a}\left(e\right)$ (with $e \in [1, ... ,N]$) from a process with a mean $\bar{\bm x}^{0}_{a}$ and covariance $\bm C^{0}_{a}$
\item \textit{Forecast}: At each time step n, update each of the N ensemble members according to PRMS forward simulation model. Then update the ensemble mean and covariance according to:
\begin{equation}
\label{eq:integration}
\bm x^{k}_{f}(e) = \mathcal{M}[\bm x^{k-1}_{f}(e), \bm b^{k}(e)] + \bm \eta^{k}(e)
\end{equation}
$$ \bar{\bm x}^{k}_{f} = \frac{1}{N}\sum_{e=1}^{N} \bm x^{k}_{f}(e)$$
$$\bm C^{k}_{ens,f} = \frac{1}{N-1}\sum_{e=1}^{N}\left(\bm x^{k}_{f}(e) - \bar{\bm x}^{k}_{f}\right)\left(\bm x^{k}_{f}(e) - \bar{\bm x}^{k}_{f}\right)^{T}$$
\item \textit{Analysis}: Assuming a measurement is obtained at time n, localize sample covariance, compute the Kalman gain, and update the network forecast:
$$\bm C^{k}_{ens,f,L} = \mathcal{L}\circ \bm C^{k}_{ens,f}$$
$$ \bm K_{ens}^{k} = \bm C^{k}_{ens,f,L} \left(\bm H^{k}\right)^{T}\left(\bm H^{k} \bm C^{k}_{ens,f,L}\left(\bm H^{k}\right)^{T} + \bm C^{k}_{obs}\right)^{-1}$$
\begin{equation}
\label{eq:analysis}
\bm x^{k}_{a}(e) = \bm x^{k}_{f}(e) + \bm K_{ens}^{k} \left(\bm y_{obs}^{k} - \bm H^{k} \bm
x^{k}_{f}(e) + \bm \chi^{k}(e)\right) 
\end{equation}
%return
\item Return to 2.
\end{enumerate}

 ~where:
\begin{enumerate}
\item[$s^{k}_{f}$ ] is the ensemble of states forecast at time k, [s x N]
\item[$\mathcal{M}$] is the model operator (PRMS) 
\item[$\bm b^{k}$] is the perturbed ensemble input, [i x N] 
\item[$\bm \eta^{k}$] is the white process noise with covariance Q, [s x N]
\item[$\bar{\bm s}^{k}_{f}$] is the ensemble sample mean of the states forecast at time k, [s]
\item[$\bm C^{k}_{ens,f}$] is the ensemble states forecast sample covariance at time k, [s x s] 
\item[$\bm H^{k}$] is the observation matrix at time k, [o x s]
\item[$\bm K_{ens}^{k}$] is the Kalman gain, [s x o]
\item[$\bm R^{k}_{obs}$] is the measurement error covariance, [o x o]
\item[$x^{k}_{a}$] is the ensemble states after analysis at time k, [s x N]
\item[$\bm y_{obs}^{k}$] is the vector of observations, [o x 1]
\item[$\bm \chi^{k}$] is the measurement perturbations with 0 mean \& covariance $\bm R^{n}_{obs}$, [o, N]
\end{enumerate}

One can refer to \cite{eve_book} for more details and for practical implementation.
%\todo{describe parameter random walk process}

%\section{Filter Validation}
%\todo{I think it means to show it works on an experiment where you know what the result should be}

%\section{Challenges}
%%% Challenges
\underline{Model Constraints}
% model constraints

The model described above has physical constraints for inputs, states, observations and parameters as described in section \ref{ssec:model_features}. During the perturbation of such quantities with normally distributed noise and after the update step of the EnKF, invalid scenarios are likely to arise. To circumvent this issue, a hard boundary check is implemented after each analysis step. Any values outside the range are set to the corresponding maximum or minimum of that state.

Major determinants of EnKF performance are the ensemble size and the error statistics.

\underline{Ensemble Size}\\
With an infinite ensemble size, the EnKF approximation will reach the true KF solution. We choose an ensemble size of 100 for practical reasons detailed in Section \ref{ssec:model_features}.

Additionally due to the small ensemble size compared to the state space size, it might be necessary to use localization when using the EnKF when spurious correlations between physically uncorrelated states (ex: geographically distant states) affect the filter performance \cite{enkf_N_cov_infl, hess-enkf_N_L}.
Distance based localization weights using linear, exponential or Gaussian decorrelation functions are typically used to attenuate the correlations between distant states and observations. 
For this study, we do not use any localization as the impact on the outflow is assumed to be negligible compared to the scheme introduced in section \ref{sec:joint_assim}.

%model error & ensemble size N
\underline{Inputs and Model Error Statistics}\\
%We assume the error from inputs (temperature, precipitation and solar radiation) and parameters are dominant in the process model and thus no state perturbations are modeled. \todo{cite} shows that such assumption is valid for hydrologic models.
The additive noise  $\bm \eta^{k}$ in the integration equation at time $k$ (\ref{eq:integration}) is modeled as a fraction of the mean of the ensemble.
This scheme was chosen because the majority of states are storage states and have zero lower bound.
Simultaneously forcing such boundaries as described above and adding daily zero mean normal noise would introduce bias to the model.
The downside of such error modeling is that the model is assumed to have perfect estimates for states with zero means.
\begin{equation}
\bm \eta_{k} \sim \mathcal{N}(\mu=0,\,\sigma=\alpha \bar{\bm x}^{k}_{f}) 
\end{equation}

 Inputs (and potentially parameters) perturbation propagates model errors that are consistent with the physics of the model, compared to only a simple daily addition of white noise to the ensemble. Inputs and observations error perturbations are modeled as independent normal distributions with standard deviations computed as the following: 

For input precipitation:
%and solar radiation distributions is computed prohttps://www.overleaf.com/15906148xtnrpdzfbvjh#portionally to their values:
\begin{equation}
\sigma_{P} = \stdPm \cdot P
\end{equation}
\begin{equation}
Pens \sim \mathcal{N}(\mu=P,\,\sigma^{2}=(0.4\cdot P)^{2})
\end{equation}

For simulated SWE measurement:
%and SWE
\begin{equation}
\sigma_{SWE} = \stdSWEm \cdot SWE
\end{equation}
%\begin{equation}
%\sigma_{swrad} = \omega \cdot swrad + \epsilon
%\end{equation}

The standard deviation for temperature is assumed to be a constant:
\begin{equation}
\sigma_{T} = \stdT ^{o}C
\end{equation}
\begin{equation}
Tens \sim T + \mathcal{N}(\mu=0,\,\sigma^{2}=2^{2})
\end{equation}

Variance inflation is required when the model forecast error is known to be underestimated. 
%In EnKF, many studies show that it is often necessary \todo{cite} to account for additional unknown errors from inputs, model and previous observation uncertainty and the analysis operation, especially when frequent assimilation updates occur. 
Without variance inflation, the filter tends to diverge and the observation fails to influence the model forecast because of the underestimated spread in the ensemble state forecast due to previous update events. Multiple studies implement variance inflation differently \cite{enkf_N_cov_infl} and some use adaptive inflation that is pre-computed at analysis step as a function of the error between the forecasted state and the observation \cite{infl_soil}.
In this study we will use a post-analysis inflation procedure with constant $\alpha_{i} = 0.9$ similar to \cite{SHI2015_Parameter_estimation}. %they used 0.5
\begin{equation}
\label{eq:post_asis_infl}
\bm{x}^{\prime}_{a, infl} = 
\left(1 - \alpha_{i}\right)\bm{x}^{\prime}_{a} + \alpha_{i}\cdot \bm{x}^{\prime}_{f}
\end{equation}
where the prime $\bm{x}^{\prime}$ denotes the ensemble anomaly $(\bm{x} - \bm{\bar{x}})$.

\section{Model Features}
\label{ssec:model_features}
As can be seen by equations in section \ref{chap:5_system_dynamics}, the model is extremely nonlinear (ex: equations \ref{eq:lwp_power4}, \ref{eq:Fsr_exp}, and \ref{eq:Fss_exp}...) and there is frequent use of thresholds to determine what situational equations are appropriate to use (ex: equation \ref{eq:snow_rain}). 
Those threshold functions also make the resulting system not continuously differentiable. 

There are some explicit constraints for model inputs, states and parameters shown in tables \ref{tab:model_inputs}, \ref{tab:model_states} and S.1. These constraints must be satisfied for the model to run and to get a physically realistic behavior.

%\underline{Model Dimension}\\
The model is spatially distributed and runs independently on each sub-basin. Each sub-basin is divided into HRUs (for surface), SSRs and GWRs (for sub-surface) spatial units. The basin-level model dimension depends on the number of these sub-units. 

To quantify the model dimension in this study, we focus on the region of interest, which is the East Branch of the North Fork of the Feather River sub-basin, highlighted in Fig. \ref{fig:feather_hrus}. 111 HRUs have been apriori selected and they are spatially co-located with the ground-layer subsurface reservoirs (SSRs) and groundwater reservoirs (GWRs). 
We will thus call the 111 regions HRUs in the remainder of this study. 
This implies that $A_{h} = A_{s} = A_{g}$ in (\ref{eq:F_basin}).
Each HRU has a total of \nStates states, shown in Table \ref{tab:model_states}.  
In our case, model parameters can be basin-wide (i.e. fixed and constant value for all HRUs) or different for each HRU. 
Some parameters are constant and others change monthly or seasonally.

The dimension of the model in the region of interest is then \the\numexpr\nStates*\nhrus\relax. 

%Note that if we wish to only estimate the state of the model, we can do so independently for each HRU given the distributed feature of the model, thus reducing the model dimension to \nStates and repeating the estimation process \nhrus times.
%However we also want to estimate global model parameters including global, mmmm what if you can estimate the global param for each HRU and then combine them?
\underline{Data}\\
In this study, model inputs are generated by various cooperators and are obtained from Pacific Gas \& Electric (PG\&E). They were estimated using the Parameter-Elevation Relationships on Independent Slopes Model (PRISM)'s historical monthly maps \cite{prism} and daily precipitation gauges. More details are available in \cite{prms_calib}.

SWE measurements are simulated by averaging for each HRU from the 90 meter resolution daily historical product in
\cite{margulis16landsat}. 

Other measurements available are the streamflow $F\_basin$ in equation \ref{eq:F_basin} downstream of the basin to check the output performance. 
The model parameters used are those obtained from the previous calibration study of the model and can be found in \cite{prms_calib}.

The average water year 2006 was chosen for simulation. Additional water years (2011, 2014) are simulated  and their results can be found in the Supplement. 2014 was a dry year while 2011 was a wet water year \cite{Climatology}.
% dry: 2014, 2003
% avg: 2006, 1985
% wet: 2011, 1993, 
We first run the model with the calibrated parameters \cite{prms_calib} which are presently used operationally without any assimilation to serve as a reference. 
Next, we proceed to assimilate the SWE state of the model in Section \ref{sec:swe_assim}.
The experiment results motivate a new joint-state parameter assimilation scheme with feedback that is explained and simulated in Section \ref{sec:joint_assim}. 

\underline{Performance Metrics}

The Root Mean Squared Error is used to compare the stream flow model output from the assimilation experiments with the measured stream flow.
\begin{equation}
RMSE = \frac{1}{T} \sum_{t=1}^{T}{\sqrt{\left(Fbasin_{meas, t} - Fbasin_{sim, t}\right)^{2}}}
\end{equation}

The RMSE of all experiments conducted are summarized in tables \ref{tab:rmse_roff} and \ref{tab:rmse_swe}.

For ensemble size, we use the maximum possible ensemble number of 100 that is practical to run on a typical user accessible computer.
The limiting factor for increasing the ensemble size in our case is the time needed to read and write state values to an ensemble of PRMS files. The PRMS used is available as a pre-compiled binaries of a Fortran code and is setup to read/write state data from/to files. 
To improve the latency of these procedures, the files were stored on RAM instead of disk and the ensemble model forecasts were generated in parallel in a multi-threaded wrapper.
Simulation time for one water year with daily assimilation is around one hour.

%2006
%\todo{insert 2006 SWE}
%\todo{insert 2006 runoff}
\section{Results}
\subsection{SWE State Assimilation}
\label{sec:swe_assim}
\begin{figure}
\begin{center}
\includegraphics[scale=1]{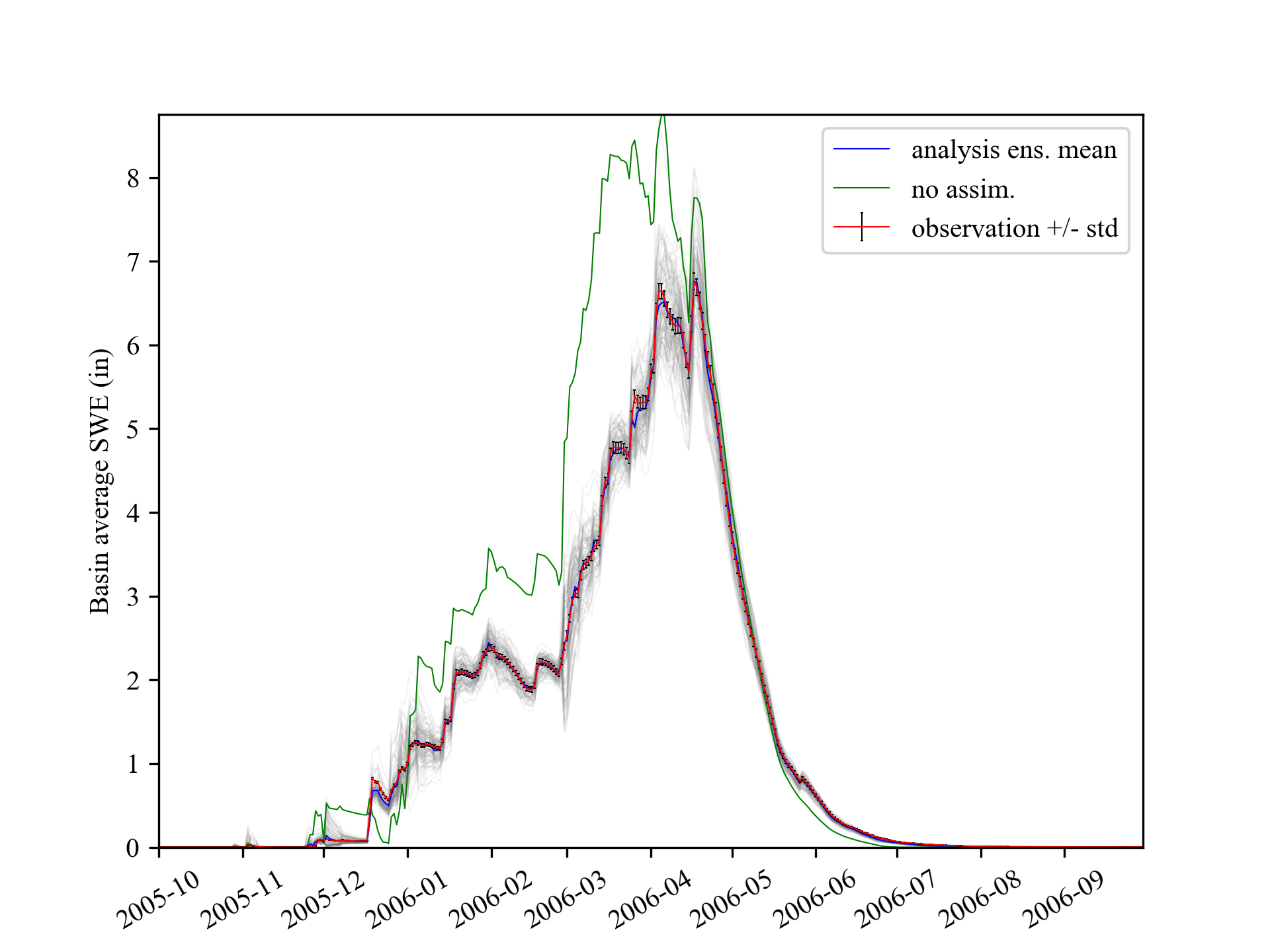}
\caption{\footnotesize{Water year 2006 basin average SWE.}}
\label{fig:2006_swe}
\end{center}
\end{figure}

%SWE follows observations
Figure \ref{fig:2006_swe} shows that the analysis SWE follows closely the ``measured" SWE.
This is expected since inflation was used to increase the uncertainty of the SWE state forecast, and thus daily observations of SWE have a relatively smaller uncertainty during the majority of the simulation days.
The remaining water year graphs are available in the Supplement.
Since the SWE observation uncertainty was modeled as a percentage of the observation value, it peaks during the peak snow season and is illustrated by the grey ensemble spread in the plot.

\begin{figure}
\begin{center}
\includegraphics[scale=1]{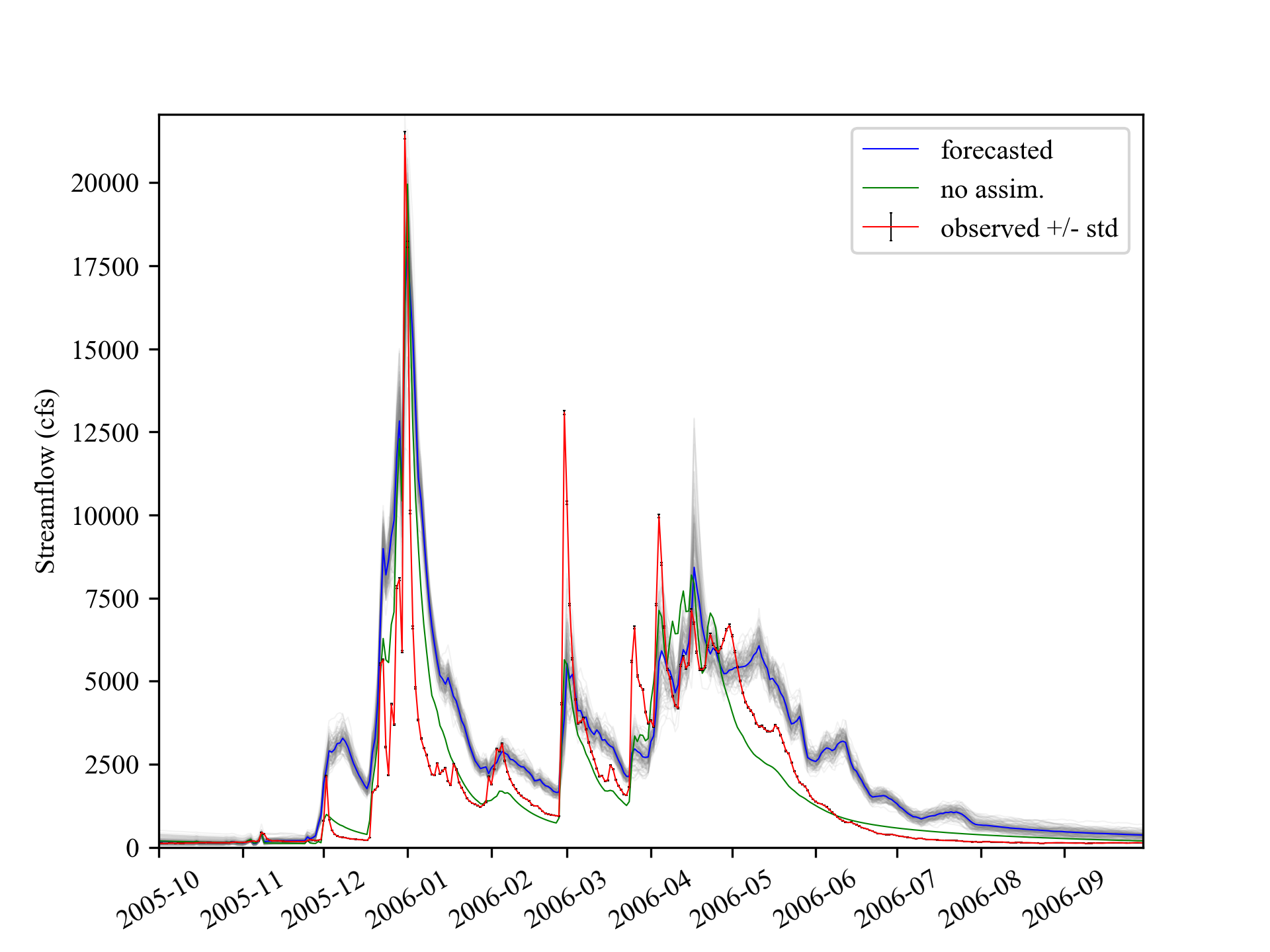}
\caption{\footnotesize{Water year 2006 basin streamflow output.}}
\label{fig:2006_roff_1}
\end{center}
\end{figure}

%Deterioration in runoff
\par Figure \ref{fig:2006_roff_1} shows a deterioration in the performance in terms of runoff output. This is consistent for all simulation years, as indicated in Table \ref{tab:rmse_roff} and in Figures S.1 and S.4 in the Supplement.
%reason_1: parameters calibrated for in/out
The main explanation for such outcome is that the model parameters used are those that were calibrated based on inputs and output only. 
The former are temperature and precipitation while the latter is the measured runoff. 
No SWE data were involved in the calibration, which implies that updating the model SWE to a more accurate estimate will not necessarily improve the model output.
%reason_2
We should also note that the SWE used as observation is not directly measured, but a modeled product. However, we assume this is not the reason of deterioration.
Furthermore, the runoff output ensemble spread is not as large as desired, since the measured runoff is not within the ensemble for the majority of the days.
%conclusion and motivation for experiment 2
\par Given that the SWE ensemble spread is relatively large, that the parameters used were calibrated without SWE knowledge, we conclude that it would be advantageous to perturb the model parameters and estimate them as well in a joint state-parameters assimilation scheme. 
Moreover a feedback consisting of the previous-day measured runoff would be necessary to estimate those parameters mostly involved in runoff generation dynamics. 

%%%%%%%%%%%%%%%%%%%%%%%%%%%%%%%%%%%%%%%%%%%%%%%%%%%%%%%%%%
%%%%%   SWE and runoff joint state-parameter assimilatiom
%%%%%%%%%%%%%%%%%%%%%%%%%%
\subsection{SWE and Runoff Joint State-Parameter Assimilation}
\label{sec:joint_assim}
%%%%%%%%%%%%%%%%%%%%%%%%%%
%%%%%   New state space
%%%%%%%%%%%%%%%%%%%%%%%%%%
\subsubsection{New State Space}
%parameters need change
\par Results of the previous experiment imply that parameters need to be altered.
Model parameters can be appended to the state vector and updated online on assimilation events in what is termed a ``Joint state-parameter estimation". 
In such method, there is no need to apriori train the model to find the optimal parameters which might not be time-independent. 
Instead, the range of each parameter must be known.
\color{\change}Parameters are indirectly updated through the cross-correlation between the  parameter and the ``state" being observed in the Kalman gain matrix.

Joint parameter-state assimilation has been used in many applications with complex models and spans different fields to name a few: assimilation of measured blood concentration into a metabolic model to update the numerous model parameters as well as other unobserved states \cite{Arnold_2014}, assimilation of measured velocity fields from GPS-equipped drifters into a shallow water model to update model parameters \cite{joint_shallow_water}, and assimilation of measured displacements into gas storage geomechanical model to reduce uncertainty of some model parameters \cite{gas}.
Results in \cite{kaipio_vadose} showed that simultaneous parameter estimation is necessary when the model perturbed by daily additive white noise is far from the truth.
\color{black}

Parameters involved in model dynamics from model states to streamflow output require streamflow measurements to be updated.
Parameters that are included in the state space were chosen based on the sensitivity results obtained from the study in \cite{sensitivity}.
%https://www.hydrol-earth-syst-sci.net/20/4655/2016/hess-20-4655-2016.pdf
We choose the parameters that score high on daily streamflow statistics of mean, CV and AR1 indicating sensitivity of parameters to each process. 
Processes chosen were snowmelt (assimilating SWE) and runoff (assimilating runoff), with emphasis on runoff.
Chosen paramters include \nLocalParAug parameters per HRU and \nGlobalParAug global parameters and are shown in Table \ref{tab:perturbed_par}. 
The state vector is augmented with these parameters.
%and the previous-day updated streamflow. %and inflated
Exponential parameters are excluded (ex: smidx\_exp) to maintain approximate Gaussian ensemble distributions, a requirement of the EnKF.

%params_pert
\begin{table}[]
\centering
\resizebox{\textwidth}{!}{%
\begin{tabular}{c|c|c|c|c|c|l}
Symbol & PRMS symbol & Quantity & Type & Range & Units & Description \\
\hline
Tms & tmax\_allsnow & 1 & constant & -10, 40 & $^{o}$F & temperature below which P is all snow \\
Tmr & tmax\_allrain & 1 & monthly & -8, 60 & $^{o}$F & temperature above which P is all rain \\
$\alpha$1 & smidx\_coef & per HRU & constant & 0.001, 0.06 & - & linear surface runoff (Fsr) coeficient \\
Asr & carea\_max & per HRU & constant & 0.0, 1.0 & - & max areal fraction contributing to surface runoff Fsr \\
$\beta$ & dday\_intcp & 1 & monthly & -60.0, 10.0 & $^{o}$day & intercept of degree-day equation \\
jc & jh\_coef & 1 & monthly & 0.005, 0.06 & /$^{o}$F & coef. used in ETp \\
Sszmax & soilmoist\_max & per HRU & constant & 0.001, 60.0 & inches & maximum soilzone water holding capacity \\
$\alpha$4 & gwflow\_coef & per GWR & constant & 0.001, 0.5 & -/day & linear coefficient routing Sgw to streamflow \\
Fzgwmax & soil2gw\_max & per HRU & constant & 0.0, 5.0 & inches & max soil excess water routed to gwStr \\
$\alpha$5 & gwsink\_coef & per GWR & constant & 0.0, 1.0 & -/day & linear coefficient for groundwater sink \\
$\alpha$2 & ssr2gw\_rate & per SSR & constant & 0.05, 0.8 & -/day & linear coefficient routing Sss to Sgw \\
$\alpha$3 & ssrcoef\_sq & per SSR & constant & 0.0, 1.0 & - & coeficient routing Sss to streamflow \\
$\beta$3 & ssrcoef\_lin & per SSR & constant & 0.0, 1.0 & -/day & linear coeficient routing Sss to streamflow\\
\hline
\end{tabular}%
}
\caption{Perturbed and updated parameters.}
\label{tab:perturbed_par}
\end{table}

\begin{figure}
\begin{center}
\includegraphics[scale=1]{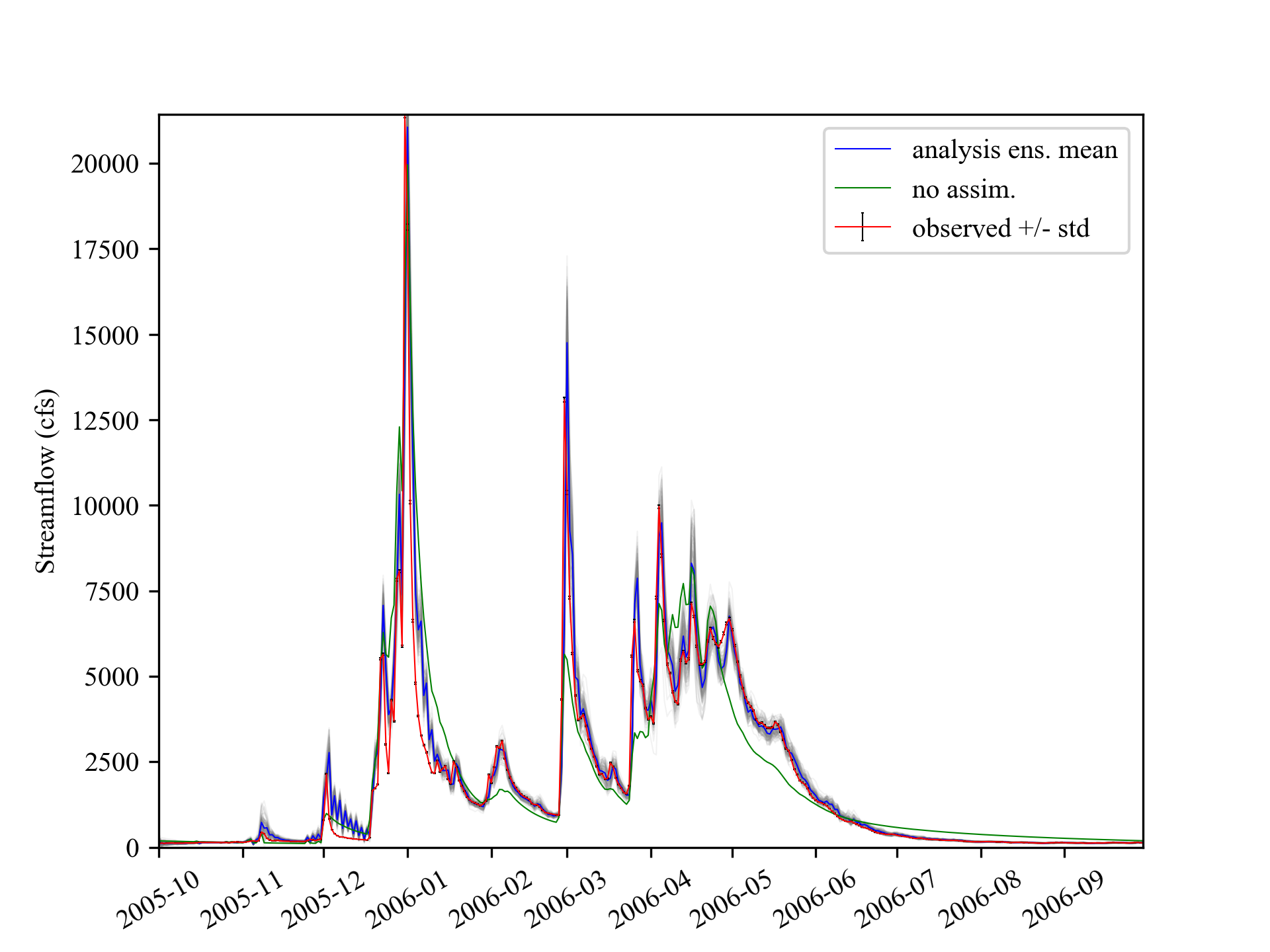}
\caption{\footnotesize{Water year 2006 basin average streamflow.}}
\label{fig:2006_roff}
\end{center}
\end{figure}

Initial parameter perturbations are modeled proportionally to  their range, such that:
\begin{equation}
\sigma_{p} = 0.25 \cdot |p_{max} - p_{min}|
\end{equation}
%Where $\kappa$'s value is \stdInitParamm initially or at the start of each water year, and 
Parameters are inflated post-analysis similarly to states as described in (\ref{eq:post_asis_infl}).
Parameters do not change during model integration, thus when the parameter ensemble nears collapse - ie. when the standard deviation becomes less than $\sigma_{target}$, inflation is performed by re-scaling the parameter ensemble perturbations to $\sigma_{target}$:
\begin{equation}
\sigma_{target} = 0.25 \cdot \sigma_{p}
\end{equation}
where $\sigma_{p}$ is the standard deviation of the initial parameter perturbation.

%observation change
Next, we additionally augment the state-space with the previous-day updated runoff, so that it can be updated on the next analysis day by the previous-day measured runoff.
The current day simulated streamflow is also appended.
Finally, the dimension of the augmented state-space becomes \the\numexpr(\nStates + \nLocalParAug)*\nhrus + \nGlobalParAug + 2\relax.

The observation vector $y_{obs}$ is augmented by the previous-day measured runoff and correspondingly, a new row is appended to the observation matrix $H$ in the analysis equation (\ref{eq:integration}).
The previous-day measured runoff is available in practical applications, because we are updating the SWE state daily.
The measured streamflow is assumed near perfect and perturbed with an independent normally distributed error with a scaling standard deviation:
\begin{equation}
\label{eq:streamflow_error}
\sigma_{F} = \stdRoffm \cdot Fbasin_{meas}
\end{equation}

%%%%%%%%%%%%%%%%%%%%%%%%%%
%%%%%   Results
%%%%%%%%%%%%%%%%%%%%%%%%%%
\subsection{Results}
As reference, we compare the results to both the estimated streamflow with no assimilation, which is the green plot shown in Figure \ref{fig:2006_roff}, and an additional estimate which is simply the previous-day measured streamflow (not visualized), also termed autoregressive AR(1) filter with equation:
\begin{equation}
Fbasin_{sim, t} = Fbasin_{meas, t-1}
\end{equation}

Results of both experiments are summarized in Tables \ref{tab:rmse_roff} and \ref{tab:rmse_swe} and show a consistent improvement in simulated streamflow compared to the no assimilation parameter-calibrated case, with up to 60\% reduction in RMSE for the wet water year 2011. 
Results were occasionally better than the naive previous-day AR(1) forecast such as the average water year 2006 (28\%), but worse during the dry water year 2014.
For all assimilation experiments, the RMSE of the basin-mean SWE state was substantially reduced ($>90\%$) as expected, since it is the state being ``measured" with $10\%$ uncertainty. 
SWE results constitute a validation of the EnKF analysis procedure.
% \usepackage{graphicx}
\begin{table}[]
\centering
\resizebox{0.8\textwidth}{!}{%
\begin{tabular}{r|r|r|r|r|r|r|r}
Water Year & no assim. & SWE update & $\Delta$(\%) & Joint SWE/Fbasin & $\Delta$(\%) & AR(1) & $\Delta$(\%) \\
\hline
2006       & 1228      & 1845       & 50           & 889                 & -28          & 1235        & 1            \\
2011       & 870       & 1891       & 117          & 360                 & -59          & 430         & -51          \\
2014       & 198       & 996        & 404          & 177                 & -10          & 126         & -36  \\

\hline
\end{tabular}%
}
\caption{Experiments performance in terms of streamflow output RMSE}
\label{tab:rmse_roff}
\end{table}

% \usepackage{graphicx}
\begin{table}[]
\centering
\resizebox{0.7\textwidth}{!}{%
\begin{tabular}{r|r|r|r|r|r}
Water Year & no assim. & SWE update & $\Delta$(\%) & Joint SWE \& Stream & $\Delta$(\%) \\ 
\hline
2006       & 1.08      & 0.02       & -98          & 0.04                & -96          \\
2011       & 1.97      & 0.05       & -97          & 0.05                & -98          \\
2014       & 0.20      & 0.01       & -93          & 0.01                & -95         \\
\hline
\end{tabular}%
}
\caption{Experiments validation in terms of basin-mean SWE RMSE.}
\label{tab:rmse_swe}
\end{table}

%%%%%%%%%%%%%%%%%%%%%%%%
% Conclusion
%%%%%%%%%%%%%%%%%%%%%%%%
\section{Conclusion}
The EnKF data assimilation framework presented succeeded in updating the modeled SWE during analysis step. SWE state assimilation alone does not improve runoff in a heavily parameterized model, where the parameters have been calibrated based on inputs/output without taking into consideration the SWE state being updated. Furthermore, results show that deterioration in accuracy can occur. We postulate SWE assimilation in a non-parametric model is likely to improve streamflow forecast accuracy.

Joint state-parameter assimilation using the previous day measured stream flow as feedback shows a substantial improvement in the accuracy of the daily estimate of streamflow (30\% reduction in RMSE for water year 2006) compared to the calibrated no-assimlation scenario, and compared to the simple previous-day AR(1) estimator (30\% reduction in RMSE for water year 2006) during average water years.

\par In reality, streamflow measurements are not as accurate as assumed in this study. It is typically computed from the river stage measurements assuming constant cross section area and other approximations that should be accounted for. Streamflow measurement uncertainty, if well known, can be modeled and accounted for in \ref{eq:streamflow_error}.
Nevertheless, what this study shows is that with a near-perfect measurement of streamflow, the EnKF framework presented improves the accuracy of the streamflow estimate. More accurate measurement methods do exist such as current meters or the ``Acoustic Doppler Current Profiler", but are more expensive.

Although the previous day estimator (AR(1)) was better for the dry year 2014, it has disadvantages over using the physical model-based assimilation framework in that it is not a reliable method, especially when the previous streamflow measurement is not available or severely inaccurate due to sensor failure or environmental events, whereas the EnKF framework would provide a more robust and physically sound estimate when measurements are missing. 
In fact, the only change required when previous day measurement of streamflow is not available is to delete one row from the observation matrix.
Moreover, the EnKF with PRMS framework is preferred to the naive AR(1) method given the forecast time window the former can potentially achieve. 
We suggest that a more sophisticated fully data-driven model that considers at least input data is required to match its potential long-term forecast accuracy such as an artificial recurrent neural network (ex: Long short-term memory, LSTMs).

We suggest as future work to replicate the study for all years where data is available.
A longer forecast window study is also of interest to stakeholders and would be a natural continuation of this report, where multiple days in the future are first simulated with model inputs from weather predictions (forecast step), after which  their streamflow outputs are updated with measurements of SWE and streamflow available at the current day (update step).
This would increase the state dimensions only by the number of days in the prediction window, keeping the assimilation tractable.
With parameters updated by measurements of streamflow and states such as SWE, current-day measurements of SWE (and potentially other states) should have a strong impact on the streamflow prediction accuracy for a long prediction window.

\section*{Acknowledgement}
\par This work was partially supported by the Civil Systems department at UC Berkeley, California Energy Commission through the grant “Improving Hydrological Snow pack Forecasting for Hydropower Generation Using Intelligent Information Systems” (EPC-14-067), Pacific Gas and Electric Co (PG\&E), the California Department of Water Resources and UC Water.

\color{\change}
I would like to thank Prof. Steven Glaser for recommending me to pursue this research work and to thank Prof. Alexandre M. Bayen for his advice and words of encouragement. 
\color{black}

Finally, special thanks to Kevin Richards (PG\&E) for sharing streamflow measurements and Feather river data pertaining to PRMS such as HRUs and calibrated parameters.

\bibliographystyle{IEEEtran}
\addcontentsline{toc}{chapter}{Bibliography}
%\ssp    % SINGLE SPACE REFERENCES (optional)
\bibliography{references.bib}

\end{document}

\maketitle

\section{Model Parameters Reference}
This section consists of Table \ref{tab:model_parameters}.

\section{Model Variables Reference}
This section consists of Table \ref{tab:model_variables}.

% \usepackage{longtable}
% Note: It may be necessary to compile the document several times to get a multi-page table to line up properly
%\scriptsize %scriptsize
%\begin{left}
%\begin{table}[]
%\centering
%\resizebox{\textwidth}{!}{
%\begin{tabular}{c|c|c|c|c|c|l}
\begin{table}[]
\centering
\resizebox{\textwidth}{!}{%
\begin{tabular}{c|c|c|c|c|c|l}
Symbol     & PRMS symbol          & Quantity & Type     & Range       & Units             & Description        \\ \hline
%\endhead

%
Pmin       & ppt\_rad\_adj        & 1        & monthly  & 0.0, 0.5    & inches            & minimum P to adjust solar radiation                             \\
$\phi$     & dday\_slope          & 1        & monthly  & 0.2, 0.9    & $^{o}$day/$^{o}$F & slope of degree-day equation                                    \\
$\beta$    & dday\_intcp          & 1        & monthly  & -60.0, 10.0 & $^{o}$day         & intercept of degree-day equation                                \\
${\gamma}s$         & radj\_sppt           & 1        & constant & 0.0, 1.0    & -                 & adjustment factor for summer solar rad with P \textgreater Pmin \\
${\gamma}w$         & radj\_wppt           & 1        & constant & 0.0, 1.0    & -                 & adjustment factor for winter solar rad with P \textgreater Pmin \\
$\lambda$2 & tmin\_lapse          & 1        & monthly  & -10.0, 10.0 & $^{o}$F/mile      & slope of minimum temperature interpolation                      \\
$\lambda$1 & tmax\_lapse          & 1        & monthly  & -10.0, 10.0 & $^{o}$F/mile      & slope of maximum temperature interpolation                      \\
$\beta$2   & tmin\_adj            & per HRU  & constant & -10.0, 10.0 & $^{o}$F           & physiographic adjustment to minimum T                           \\
$\beta$1   & tmax\_adj            & per HRU  & constant & -10.0, 10.0 & $^{o}$F           & physiographic adjustment to maximum T                           \\
ds         & covden\_sum          & per HRU  & constant & 0.0, 1.0    & -                 & vegetation areal coverage during summer                         \\
dw         & covden\_win          & per HRU  & constant & 0.0, 1.0    & -                 & vegetation areal coverage during winter                         \\
Tms        & tmax\_allsnow        & 1        & monthly  & -10.0, 40.0 & $^{o}$F           & Tmax below which all P is snow                                  \\
Tmr        & tmax\_allrain        & 1        & constant & -8.0, 60.0  & $^{o}$F           & Tmax above which all P is rain                                  \\
$\zeta$          & adjmix\_rain         & 1        & monthly  & 0.6, 1.4    & -                 & rain fraction of mixed rain-snow P                              \\
Crs       & srain\_intcp         & per HRU  & constant & 0.0, 1.0    & inches            & canopy summer rain interception storage capacity                \\
Crw       & wrain\_intcp         & per HRU  & constant & 0.0, 1.0    & inches            & canopy winter rain interception storage capacity                \\
Cs        & snow\_intcp          & per HRU  & constant & 0.0, 1.0    & inches            & canopy winter snow interception storage capacity                \\
jc         & jh\_coef             & 1        & monthly  & 0.005, 0.06 & /$^{o}$F          & coef. used in ETp                                               \\
$\alpha$   & albedo               & per HRU  & constant & 0.0, 1.0    & -                 & fraction of radiation reflected by snow                         \\
$\psi$     & rad\_trncf           & per HRU  & constant & 0.0, 1.0    & -                 & transmission coefficient for Rsw through winter canopy          \\
$\epsilon$ & emis\_noppt          & 1        & constant & 0.757, 1.0  & -                 & emissivity of air with no P                                     \\
$\omega$   & cecn\_coef           & 1        & monthly  & 2.0, 10.0   & cal/$^{o}$C       & convection \& condensation coefficient                          \\
$\xi$          & potet\_sublim        & 1        & constant & 0.0, 1.0    & -                 & fraction of ETp sublimated                                      \\
$\rho$init & den\_init            & 1        & constant & 0.01, 0.5   & g/cm$^{3}$             & density of new snow P                                           \\
$\rho$max  & den\_max             & 1        & constant & 0.1, 0.8    & g/cm$^{3}$             & maximum average snowpack density                                \\
$\tau$          & settle\_const        & 1        & constant & 0.01, 0.5   & -                 & settlement time constant for snowpack                           \\
fi         & hru\_percent\_imperv & per HRU  & constant & 0.0, 0.999  & -                 & HRU impervious fraction                                         \\
Simax &	imperv_stor_max	& per HRU	& constant	& 0.0, 0.1	& inches	& maximum Simp	\\												
Sszmax     & soilmoist\_max       & per HRU  & constant & 0.001, 60.0 & inches            & maximum soilzone water holding capacity                         \\
$\alpha$1  & smidx\_coef          & per HRU  & constant & 0.001, 0.06 & -                 & linear surface runoff (Fsr) coeficient                          \\
$\theta$1  & smidx\_exp           & per HRU  & constant & 0.1, 0.5    & 1/inch            & exponential surface runoff coeficient                           \\
$\alpha$3  & ssrcoef\_sq          & per SSR  & constant & 0.0, 1.0    & -                 & coeficient routing Sss to streamflow                            \\
$\beta$3   & ssrcoef\_lin         & per SSR  & constant & 0.0, 1.0    & -/day             & linear coeficient routing Sss to streamflow                     \\
$\alpha$2  & ssr2gw\_rate         & per SSR  & constant & 0.05, 0.8   & -/day             & linear coefficient routing Sss to Sgw                           \\
$\theta$2  & ssr2gw\_exp          & per SSR  & constant & 0.0, 3.0    & -                 & exponential coefficient routing Sss to Sgw                      \\
smax       & ssrmax\_coef         & per SSR  & constant & 1.0, 20.0   & inches            & coefficient routing Sss to Sgw                                  \\
$\alpha$4  & gwflow\_coef         & per GWR  & constant & 0.001, 0.5  & -/day             & linear coefficient routing Sgw to streamflow                    \\
$\alpha$5  & gwsink\_coef         & per GWR  & constant & 0.0, 1.0    & -/day             & linear coefficient for groundwater sink                         \\
Fzgwmax    & soil2gw\_max         & per HRU  & constant & 0.0, 5.0    & inches            & max soil excess water routed to gwStr                           \\
Acurve     & snarea\_curve        & 11       & constant & 0.0, 1.0    & -                 & snow area depletion curve: fsca vs SWEmax                       \\
SWEmax     & snarea\_threah       & per HRU  & constant & 0.0, 200.0  & inches            & max SWE for each HRU below which snow patch occurs              \\
Asr        & carea\_max           & per HRU  & constant & 0.0, 1.0    & -                 & max areal fraction contributing to surface runoff Fsr  \\
\hline
\end{tabular}%
}
\caption{Model parameters}
\label{tab:model_parameters}
\end{table}
%\end{left}

% \usepackage{longtable}
% Note: It may be necessary to compile the document several times to get a multi-page table to line up properly
\begin{table}[]
\centering
\resizebox{0.8\textwidth}{!}{%
\begin{tabular}{c|c|c|c|l}
Symbol    & PRMS symbol     & Quantity & Units     & Description  \\   \hline 
%\endhead
%
Rsw       & swrad           & per HRU  & Langleys  & shortwave radiation                                \\
r         & solf            & per HRU  & -         & ratio of actual to potential daily solar radiation \\
Pn        & ppt\_net        & per HRU  & inches    & total net precipitation                            \\
Pt        & throughfall     & per HRU  & inches    & precipitation throughfall after interception       \\
Sc        & AvailCanSt      & per HRU  & acre-inch & available storage in canopy                        \\
ETp       & potet           & per HRU  & inches    & potential evapotranspiration                       \\
ETpa      & apet            & per HRU  & inches    & available potential evapotranspiration             \\
$\Delta$E & tcal            & per HRU  & Langleys  & net snowpack energy balance                        \\
I         & cal             & per HRU  & Langleys  & total incoming longwave                            \\
Qv        & cecsub          & per HRU  & Langleys  & convection and latent heat from condensation       \\
Rn        & swn             & per HRU  & Langleys  & net shortwave radiation                            \\
Ip        & lwp             & per HRU  & Langleys  & perfect black-body emission                        \\
Tavg      & temp            & per HRU  & $^{o}$F   & average temperatue                                 \\
Qc        & qcond           & per HRU  & Langleys  & conducted heat                                     \\
B         & sub             & per HRU  & inches    & sublimation                                        \\
Pns       & net\_snow       & per HRU  & inches    & snow portion of Pn                                 \\
D         & pk\_depth       & per HRU  & inches    & snowdepth                                          \\
M         & snowmelt        & per HRU  & inches    & snowmelt                                           \\
fsa       & snowcov\_area   & perHRU   & .         & fractional snow covered area                       \\
Pnr       & net\_rain       & per HRU  & inches    & rain portion of Pn                                 \\
Psri      & hru\_sroffi     & per HRU  & inches    & impervious surface runoff                          \\
ei        & hru\_impervevap & per HRU  & inches    & imprevious region evaporation                      \\ \hline                    
\end{tabular}%
}
\caption{Model variables}
\label{tab:model_variables}
\end{table}

\section{Water Year 2011 Results}
\subsection{SWE update}
\begin{figure}
\begin{center}
\includegraphics[scale=1]{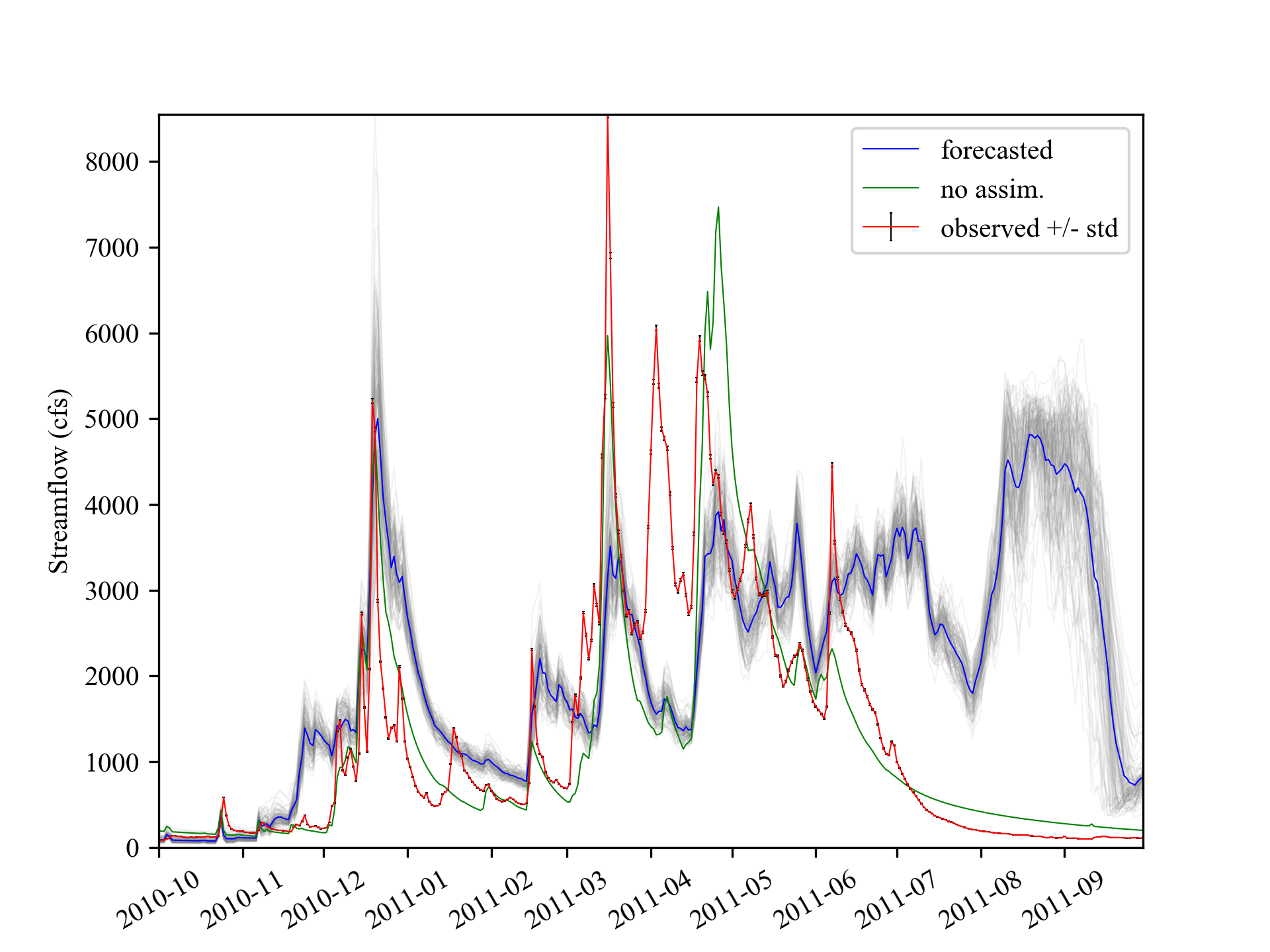}
\caption{\footnotesize{Water year 2011 basin average streamflow (SWE update).}}
\label{fig:2011_roff_1}
\end{center}
\end{figure}

This section consists of Figure \ref{fig:2011_roff_1}.

\subsection{Joint update}
\begin{figure}
\begin{center}
\includegraphics[scale=1]{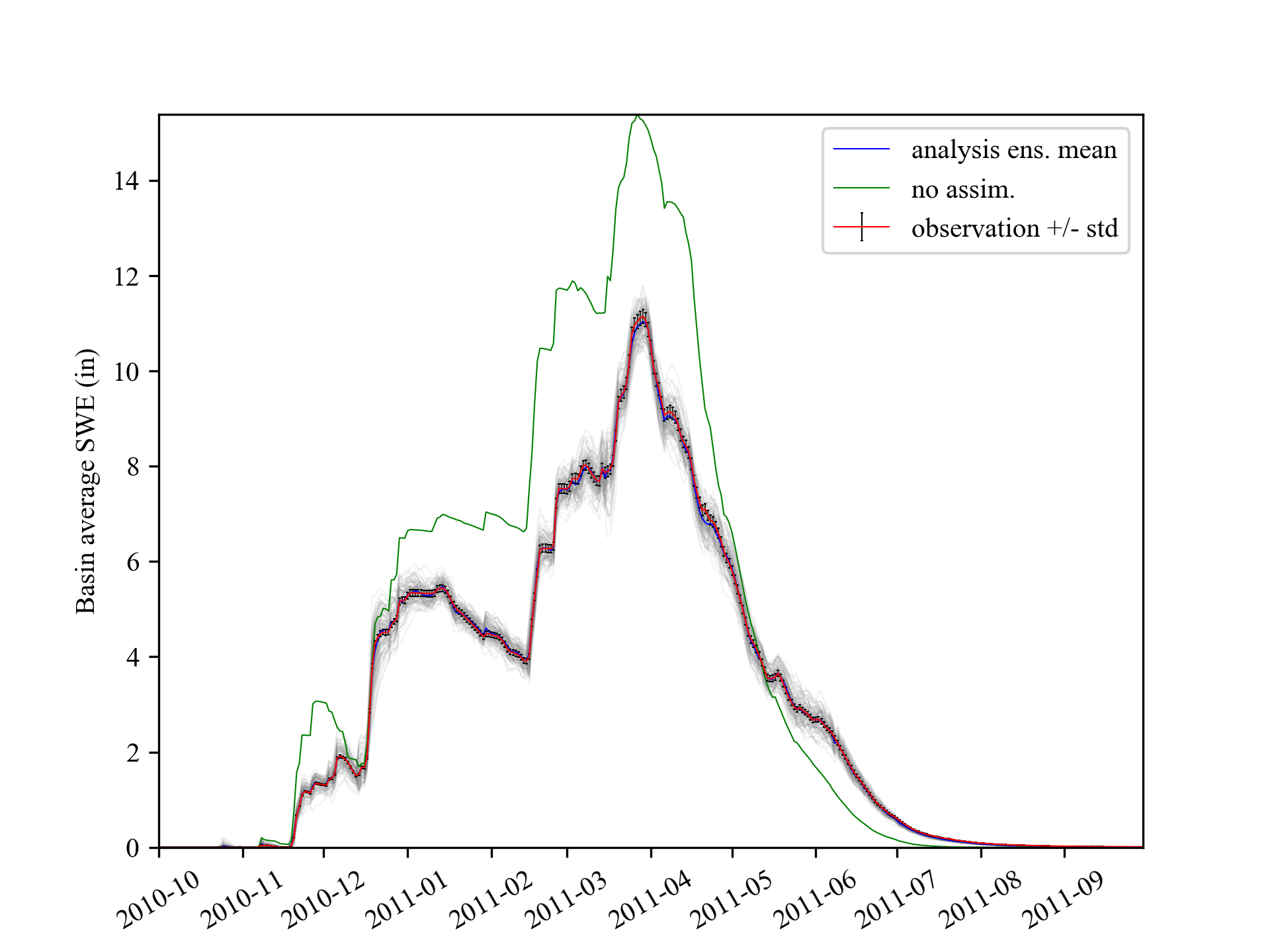}
\caption{\footnotesize{Water year 2011 basin average SWE. (SWE or Joint update)}}
\label{fig:2011_swe}
\end{center}
\end{figure}

\begin{figure}
\begin{center}
\includegraphics[scale=1]{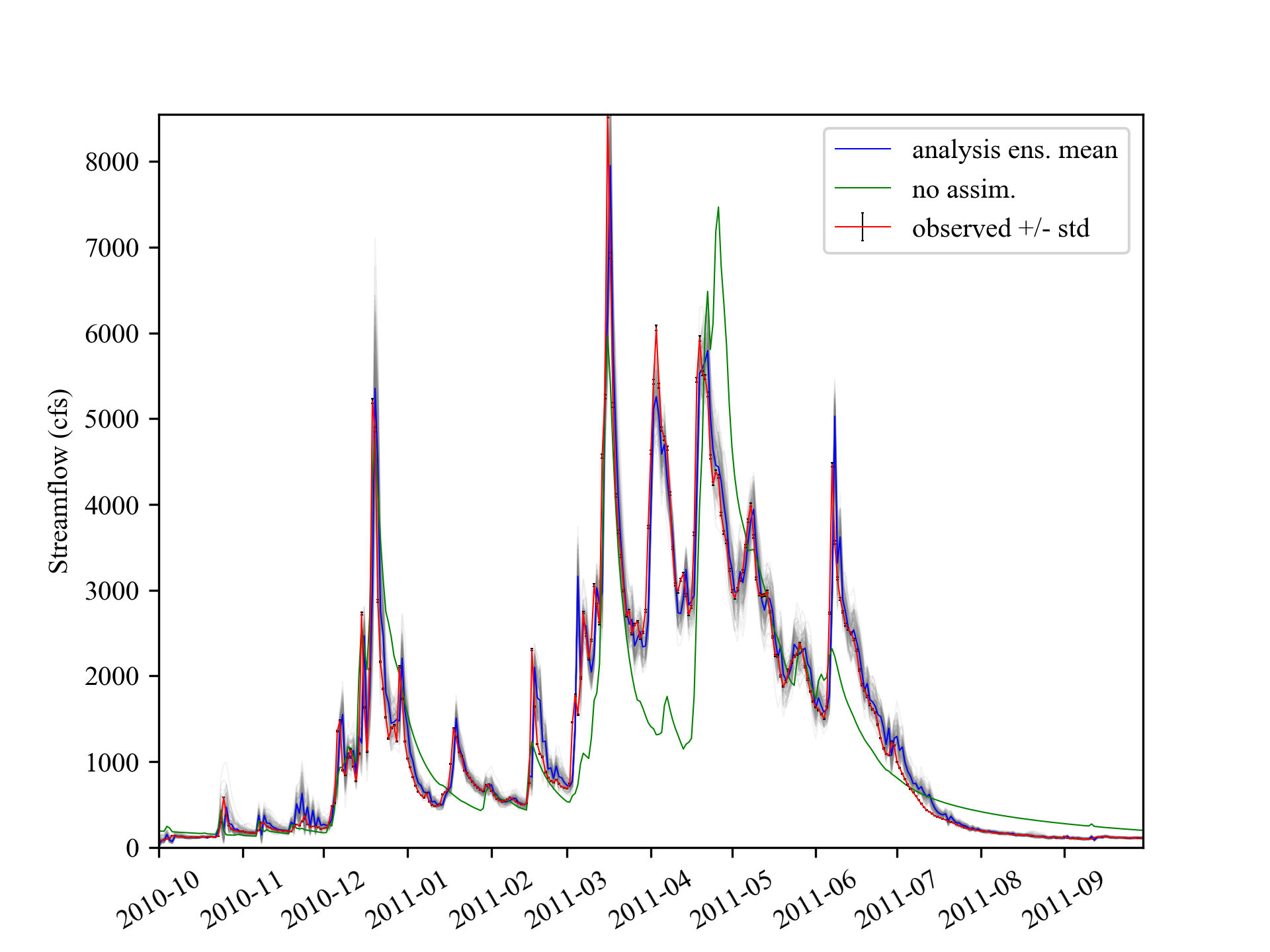}
\caption{\footnotesize{Water year 2011 basin average streamflow (Joint update).}}
\label{fig:2011_roff}
\end{center}
\end{figure}

This section consists of Figures \ref{fig:2011_swe} and \ref{fig:2011_roff}.

\section{Water Year 2014 Results}
\subsection{SWE update}
\begin{figure}
\begin{center}
\includegraphics[scale=1]{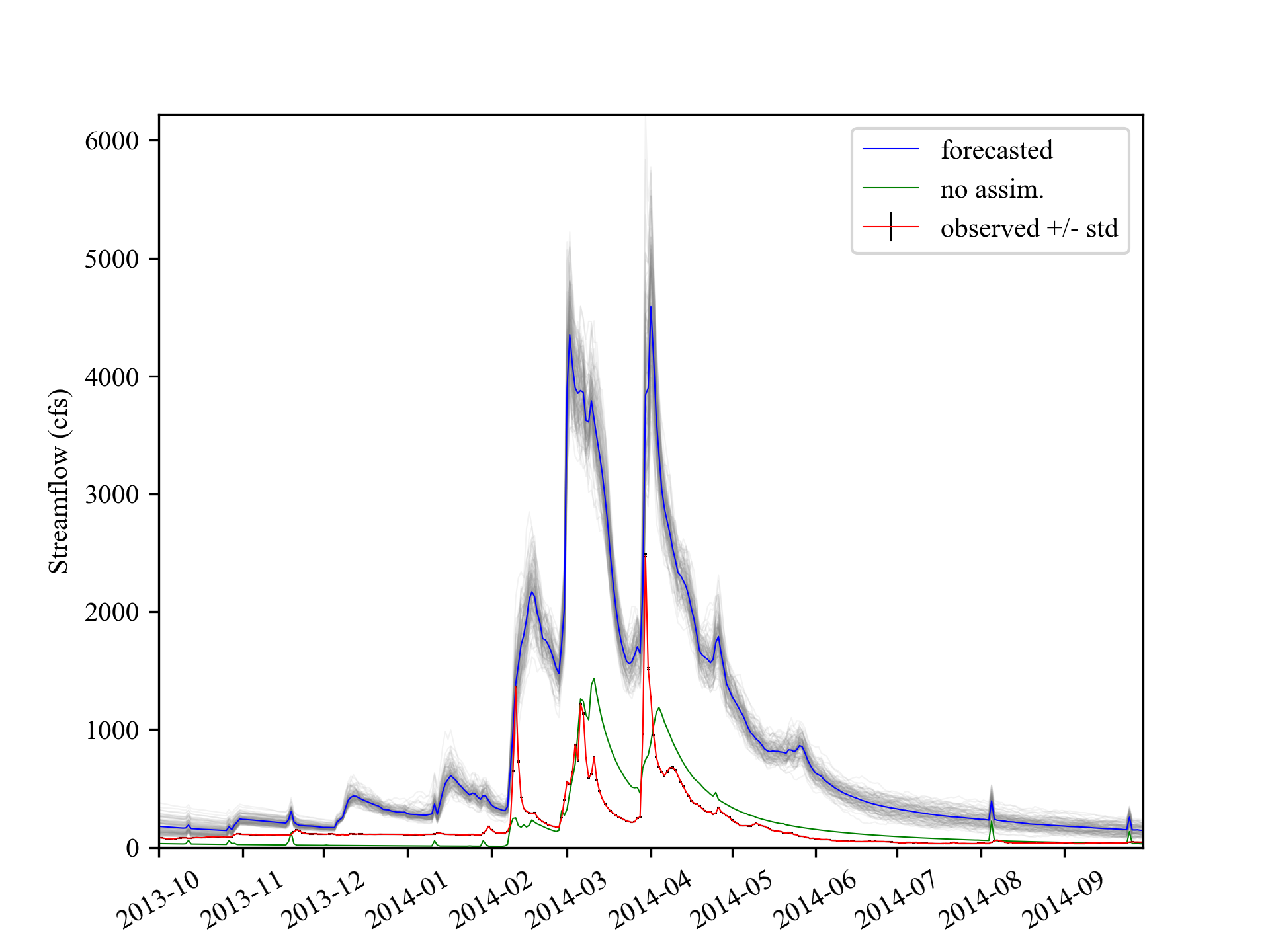}
\caption{\footnotesize{Water year 2014 basin average streamflow (SWE update).}}
\label{fig:2014_roff_1}
\end{center}
\end{figure}

This section consists of Figure \ref{fig:2014_roff_1}.

\subsection{Joint update}
\begin{figure}
\begin{center}
\includegraphics[scale=1]{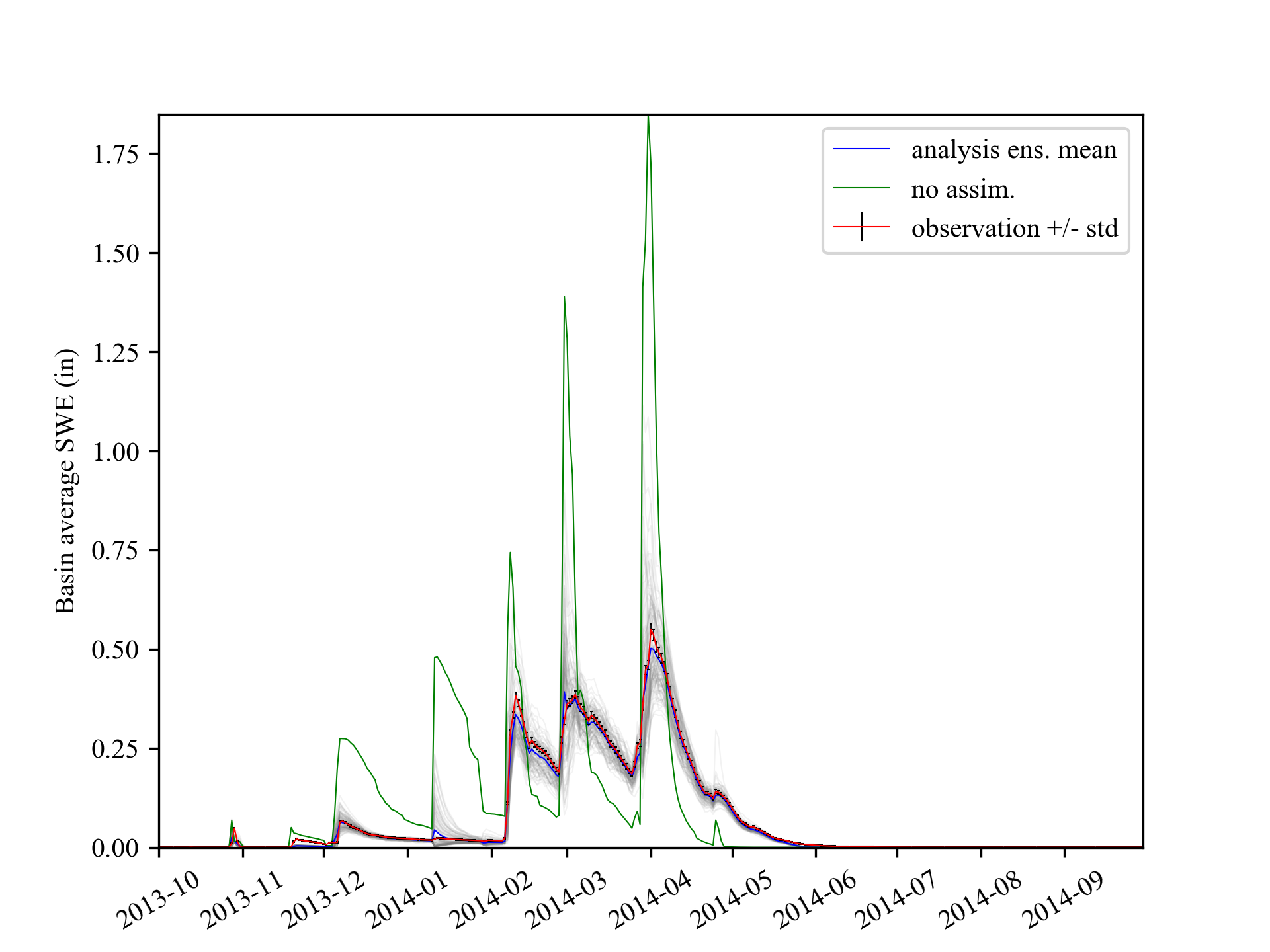}
\caption{\footnotesize{Water year 2014 basin average SWE. (SWE or Joint update)}}
\label{fig:2014_swe}
\end{center}
\end{figure}

\begin{figure}
\begin{center}
\includegraphics[scale=1]{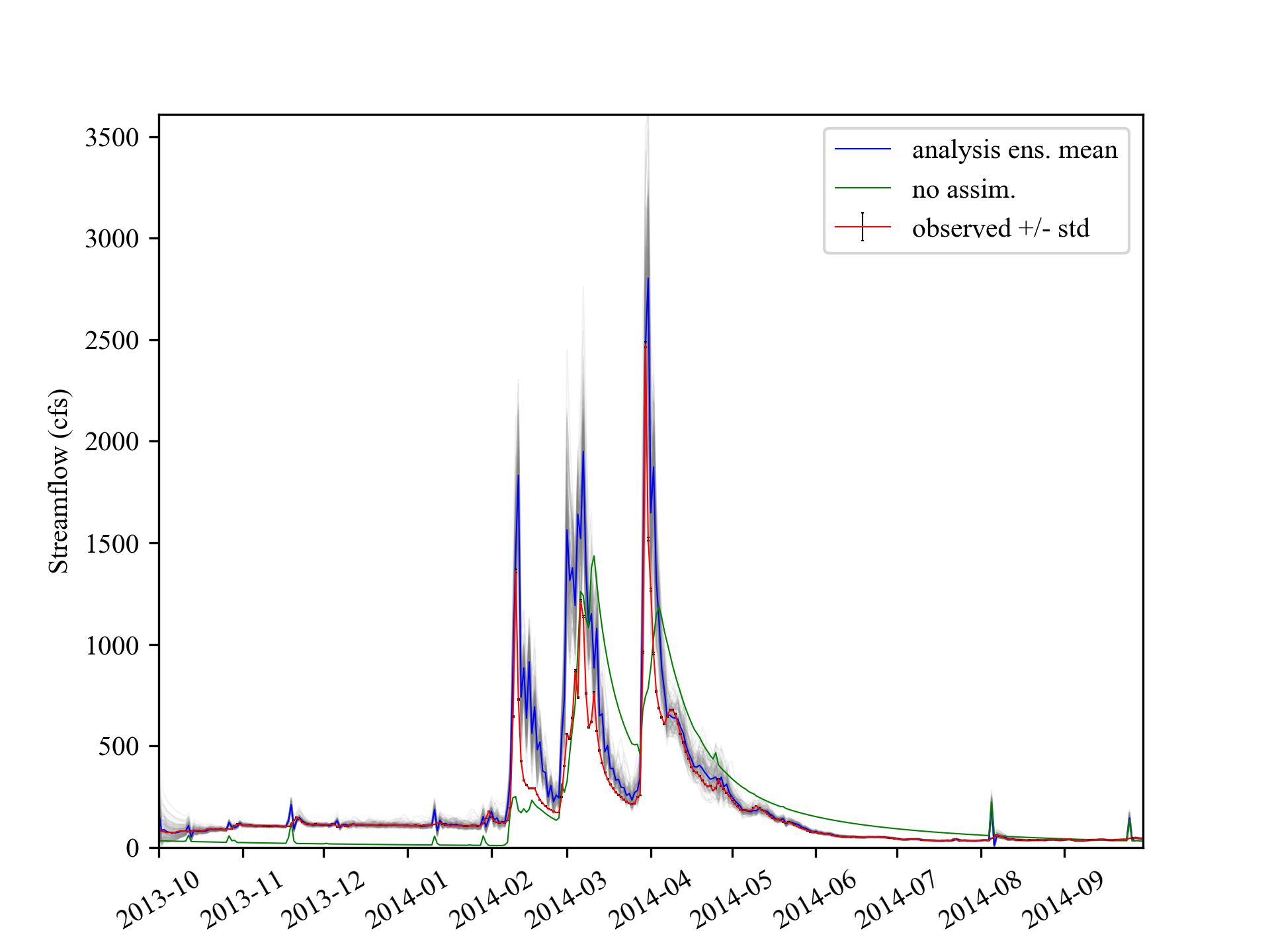}
\caption{\footnotesize{Water year 2014 basin average streamflow. (Joint update)}}
\label{fig:2014_roff}
\end{center}
\end{figure}

This section consists of Figures \ref{fig:2014_swe} and \ref{fig:2014_roff}.

